\documentclass[pra,floatfix,amsmath,superscriptaddress,twocolumn]{revtex4}
\usepackage{amssymb}
\usepackage{graphicx}
\usepackage{graphics}
\usepackage{amsmath}
\usepackage{amsthm}
\usepackage{color}

\def\sz{\sigma_z}

\def\x{X}

\def\i{{\bf i}}

\def\j{{\bf j}}

\def\k{{\bf k}}

\def\rhosigma{\sigma_S}

\newcommand{\bea}{\begin{eqnarray}}
\newcommand{\eea}{\end{eqnarray}}
\def\bi{\begin{itemize}}
\def\ei{\end{itemize}}
\def\bc{\begin{center}}
\def\ec{\end{center}}
\def\E{{\cal E}}

\def\C{\hbox{$\mit I$\kern-.7em$\mit C$}}
\def\R{\hbox{$\mit I$\kern-.6em$\mit R$}}

\def\ket#1{|#1\rangle}
\newcommand{\one}{\mbox{$1 \hspace{-1.0mm}  {\bf l}$}}
\def\tr{\mathrm{tr}}
\def\ket#1{\left| #1\right>}
\def\bra#1{\left< #1\right|}

\newcommand{\proj}[1]{\ket{#1}\bra{#1}}

\newtheorem{theorem}{Theorem}

\begin{document}

\title{Decoherence of many--body systems due to many--body interactions}

\author{T. Carle}
\affiliation{Institute for Theoretical Physics, University of
Innsbruck, Innsbruck, Austria}
\author{H. J. Briegel}
\affiliation{Institute for Theoretical Physics, University of
Innsbruck, Innsbruck, Austria}
\affiliation{Institute for Quantum Optics and Quantum Information, Austrian Academy of Science, Innsbruck, Austria}
\author{B. Kraus}
\affiliation{Institute for Theoretical Physics, University of
Innsbruck, Innsbruck, Austria}

\begin{abstract}
We study a spin-gas model, where  $N_S$ system qubits are interacting with $N_B$ bath qubits via many--body interactions. We consider multipartite Ising interactions and show how the effect of decoherence depends on the specific coupling between the system and its environment. For instance, we analyze the influence of decohenerce induced by $k$--body interactions for different values of $k$. Moreover, we study how the effect of decoherence depends on the correlation between baths that are coupled to different individual system qubit and compare Markovian with non--Markovian interactions. As examples we consider specific quantum many--body states and investigate their evolution under several different decoherence models. As a complementary investigation we study how the coupling to the environment can be employed to generate a desired multipartite state.
\end{abstract}
\maketitle

\section{Introduction}
In any realistic scenario the system of interest is never completely decoupled from its environment. This coupling causes the system to decohere which is believed to be the main reason why macroscopic quantum systems do not occur in nature. The effect of decoherence has widely been investigated in the literature. There, different decoherence models have been assumed based on interactions with a harmonic oscillator bath \cite{LeCh87}, a spin bath \cite{PrSt00} or a spin gas \cite{CaBr}.

We will consider here a microscopic decoherence model, where both, the system and the environment are described by qubits.
For a general spin model, where $N_S$ system ($S$) qubits are coupled to $N_B$ bath ($B$) qubits, the interaction would be described by a general unitary, $U_{SB}$, acting on the whole system. Assuming that the initial state of the bath would be described by the state $\rho_B$, the evolution of the system qubits would be governed by the completely positive map ${\cal E}(\rho_S)=\tr_B(U_{SB} \rho_S \otimes \rho_B U_{SB}^\dagger)$. Due to the exponential scale of the dimension of the Hilbert space the investigation of the evolution of the system qubits under such a general map is unfeasible. Thus, one has to restrict the coupling $U_{SB}$ to a certain class of unitaries and therefore to a certain class of coupling Hamiltonians $H_{SB}$. A reasonable choice is to consider only generalized Ising interactions, which describe the effect of multi-qubit collisions.
The coupling Hamiltonian for this choice $H_{SB}$ is of the form $H_{SB}=\sum_{\bf i} \alpha_{\bf i} \sigma^{\bf i}_z$, with ${\bf i}$ denoting a $(N_S+N_B)$--bit string. Each term in the Hamiltonian corresponds to a certain collision. The unitary $U_{SB}=e^{-iH_{SB} t}$, which governs the evolution of the state, is a phase gate, i.e. a unitary which is diagonal in the computational basis. Due to the fact that all terms in the Hamiltonian commute, $U_{SB}$ can be written as a product of phase gates, where each factor corresponds to a certain collision. The order of the phase gates, i.e. the number of qubits on which the phase gates acts non--trivially, is equivalent to the number of qubits which collided. The value of the phase in the phase gate depends on the interaction strength. In this article we will describe the evolution of the system in terms of phase gates for arbitrary phases. That is, we will not consider the evolution as a function of time, but will rather assume that due to a certain physical interaction a specific phase gate, $U_{SB}$ is given. The investigation of such a general model is only feasible because all the phase gates corresponding to different collisions commute.

In this article we investigate first the effect of decoherence induced by multi-qubit Ising interactions. Second, we will show how the coupling between the environment and the system can be employed to prepare the system qubits in a desired state. The first part of this article constitutes a generalization of the work presented in \cite{HartDu}, where the decoherence of a spin gas has been investigated. In contrast to \cite{HartDu} we do not restrict ourself to bipartite interactions, but consider multipartite interactions and the resulting decoherence. Obviously, considering many--qubit phase gates leads to a much bigger variety of different interactions. For two--qubit phase gates, the only relevant interaction is the one where one bath qubit is interacting with one system qubit. In the case of three qubit gates, we could for instance consider 3--qubit interactions between either $2$ system qubits and $1$ bath qubit or between $1$ system qubit and $2$ bath qubits. Apart from that, considering several system qubits, which are interacting for instance with the bath qubits via a three--qubit phase gate, the bath qubits could either be independent, or not. In this case, non, one, or both of the bath qubits could interact with both system qubits. The effect of decoherence will strongly depend on the correlations within the bath. Considering more general phase gates (acting non--trivially on $k$ systems) allows for correspondingly more different kind of interactions.

The aim of this paper is to get a better understanding of the effect of non--local decoherence on a multipartite system.  This will be achieved by comparing the influence of decoherence for different interaction models. For instance, we will compare the effect of two--qubit phase gates to the one of three and more qubit gates. Moreover, we will compare Markovian to non--Markovian processes. We will furthermore investigate how the correlation in the bath qubits affect the decoherence. Even though we will consider the most general dephasing maps, our main focus will be on those maps, where each system qubit is only interacting with bath qubits, i.e. there is no interaction between two, or more system qubits and the bath. We will call these maps purely dephasing maps. The reason for focusing on purely dephasing maps is that they correspond to the realistic situation where the processes in which several system qubits collide with bath qubits can be ignored. Another reason for focusing on purely dephasing maps is that they cannot generate entanglement. Thus, they are perfectly suited to investigate the effect of decoherence on the entanglement and coherence of the system qubits.

The outline of the paper is the following. First of all we will review the notion of a certain class of multipartite states, called locally maximally entangleable (LME) states. As we will see, the reduced states of LME states (LMESs) play an important role in our considerations of dephasing maps. The projected--entangled--pair state (PEPS) picture will be used to compute their reduced states. In Sec. \ref{SecDecMap} we will consider the most general dephasing maps and will derive a simple relation between the maps and the corresponding LMES. After that, we will focus (apart from Sec. \ref{SecLMEgen}) on purely dephasing maps which result from interactions where each system qubit is only coupled to baths qubits and not to other system qubits. We will show that all these maps are separable, i.e. they can be written as a convex combination of local maps and will express them in terms of simple local maps. In order to study how the bath--correlation influences the decoherence, we will derive a simple expression of the maps which describe the following scenario. We consider $m$ bath qubits which are all interacting with each system qubit. Each of the system qubits is also interacting with some other bath qubits, which are coupled to the $m$ bath qubits, but not to any other system qubit. The influence of correlated noise will be investigated by considering the maps described above for different values of $m$. In Sec. \ref{SecComparison} we will analyze how the effect of decoherence depends on the various coupling scenarios. In particular, we will investigate the Markovian case and compare the effect of decoherence to the non--Markovian case for a general input state (Sec. \ref{SecMarkovian}). Furthermore, we will compare the effect of phase gates of order $k$ for different values of $k$. As examples we will determine the effect of decoherence on arbitrary single system qubit state, two system qubits states and certain multipartite states (Sec. \ref{SecDiffInter}).
We will show that the intuition, that the entanglement shared by the system qubits is less disturbed, if the baths of the system qubits are stronger correlated, that is $m$ is large, is not true. It will be demonstrated that for certain multipartite states the entanglement is more robust under the decoherence of independent baths than under correlated ones.
In Sec. \ref{SecLMEgen} we will consider the complementary process to the one considered above. There we will employ the interaction between the system and bath qubits in order to prepare the system qubits in a desired multipartite state. In particular, we will show how LMESs can be generated via unitary interactions between the system qubits and the bath qubits.

Through out the paper we will use the following notation. The subscript of an operator will always denote the
system it is acting on, or the system it is describing. We denote by
${\bf i}$ the classical bit--string $(i_1,\ldots, i_n)$ with
$i_k\in\{0,1\}$ $\forall k\in \{1,\ldots, n\}$ and $\ket{{\bf
i}}\equiv\ket{i_1,\ldots, i_n}$ denotes the computational basis.
Normalization factors as well as the tensor product symbol will be
omitted whenever it does not cause any confusion and $\one$ will
denote the normalized identity operator.

\section{Locally maximally entangleable states}

Recently a new classification of
multipartite states has been proposed in order to study the properties of pure quantum states describing
$n$ qubits \cite{KrKr08}. There, the following question has been addressed. For which
states, $\ket{\Psi}$, do there exist local controlled operations, $
C_l=\sum_i U_l^i \otimes \ket{i}_{l_a}\bra{i},$ where $U_l^i$ are
unitary operations acting on system $l$ and $\ket{i}_{l_a}\bra{i}$
is acting on the auxiliary system attached to system $l$, such that
the state $C_1 \otimes C_2 \otimes \ldots \otimes C_n
\ket{\Psi}_S(\ket{0}+\ket{1})_A^{\otimes n}$ is maximally entangled
between the system ($S$) and the auxiliary system ($A$). States for which this is possible were called {\it Locally Maximally
entangleable States} (LMESs). Important examples of these states are all stabilizer states. In \cite{KrKr08} it has been shown that any LMES is (up to local unitary operations) of the form

\bea \label{LME}
\ket{\Psi}=\sqrt{\frac{1}{2^{n}}}\sum_{{\bf i}} e^{i \alpha_{{\bf
i}}}\ket{{\bf i}}\equiv U^\Psi_{ph}\ket{+}^{\otimes n},\eea where
$\alpha_{\bf{i}}\in \R$ and $U^\Psi_{ph}$ denotes the diagonal
unitary operator with entries $e^{i\alpha_{{\bf i}}}$ (phase gate). Note that the $2^n$ real phases $\alpha_{\bf{i}}$ characterize the LMES (up to local unitaries). From Eq (\ref{LME}) it can be easily seen that any LMES can be prepared by applying generalized phase gates, acting on up to all qubits, to a
product state, i.e. non--entangled pure state. That is, any
LMES $\ket{\Psi}$ can be written as
\bea \label{Prep} \ket{\Psi}=U_{1,\ldots, n} \prod
U_{i_{k_1},\ldots, i_{k_{n-1}}}\cdots \prod U_i\ket{+}^n,
\eea
where $U_{i_{k_1},\ldots, i_{k_l}}$ is a phase gate acting non--trivially on $l$ qubits. For instance, $U_{123}=\one-(1-e^{i\phi_{123}})\proj{111}_{123}$, with $\phi_{123}\in \R$ is a three--qubit phase gate. It is straightforward to see that in
this hierarchical way the $2^n$ phases $\alpha_{{\bf i}}$ can be generated. Thus, any LMES can be prepared using generalized phase
gates, which could result from a generalized Ising interaction. If
$\alpha({\bf i})$ is a polynomial of degree $k$ (as a function of
${\bf i}=(i_1,\ldots, i_n)$) then the corresponding state can be
prepared using only $k$--body interactions. Thus, the correlations in the coefficients
are directly related to a preparation scheme and therefore to the
entanglement contained in the state.

Apart from those properties, LMESs have the following properties \cite{KrKr08}: (i)
According to their definition, LMESs are characterized by the fact
that their global information can be washed out by maximally
entangling the system qubits to local auxiliary qubits. (ii) A
state is LME iff it can be used to encode locally the maximum
amount of $n$ independent bits. (iii) For any LMES, $\ket{\Psi}$, one can construct a complete set of commuting
unitary observables such that $\ket{\Psi}$ is the unique eigenstate
with eigenvalue one for all these observables (the so--called
generalized stabilizer, see also Sec. \ref{SecLMEgen}). (iv) LMESs can have an exponentially large
quantum Kolmogorov complexity \cite{MoBr06}. That is, for certain LMESs it requires exponentially many classical bits to describe the state.

Note that this classification of multipartite states is interesting for several quantum information theoretic tasks.
First of all, (i) is equivalent to the fact that all the
information contained in a LMES can be coherently "copied"
into local auxiliary systems by local unitary operations. This
seems to be a crucial property shared by those states which are
useful for quantum information tasks, like one--way quantum
computing, quantum error correction and quantum secret sharing.
Second, due to the preparation scheme [Eq. (\ref{Prep})], LMES can be entangled
in many different, but hierarchical ways. Product states but also
stabilizer states are all LME. Stabilizer states or more generally,
weighted graph states \cite{CaBr, HartDu, HeBr} are those LMES which require
only two--qubit phase gates for their preparation. LMES, however,
can be much more entangled than the stabilizer states. Moreover,
there exist, in contrast to stabilizer states, complex LMES [(iv)],
which are necessarily highly entangled \cite{MoBr06}.

In the following subsection we will consider the reduced state of LMESs, which will be relevant for our considerations of dephasing maps. In Sec. \ref{SecLMEgen} we will show how LMESs can be generated using either dissipation or unitary interactions.

\subsection{Reduced state of LMESs}
\label{redStates}

In Sec. \ref{SecDecMap} we will show that dephasing maps are strongly related to the reduced states of LMESs. Due to this fact we will present here some methods for computing them.

Let us consider an arbitrary $n$--qubit LMES, $\ket{\Psi}=U\ket{+}^{\otimes n}=\sum_{\i} e^{i \alpha_\i}\ket{\i}$, where $\i=i_1,\ldots, i_n$ and $U$ denotes an arbitrary phase gate. We call a phase gate, $U$, pure phase gate of order $k$ if it can be written as $U=\one+(e^{i\phi}-1)\proj{1}^{\otimes k}$. A phase gate will be said to be of order $k$ if it can be decomposed into pure phase gates which are maximally of order $k$. Note that any phase gate can be decomposed into pure phase gates. We will say that a qubit is overlapping if there exist at least two pure phase gates in the decomposition of $U$, which act non--trivially on this qubit. If there are two or more qubits, say qubit $1$ and $2$, for which there exists more than one pure phase gate, which act on them simultaneously, we say that the interaction possesses an overlapping edge between qubit $1$ and $2$.

Let us now consider an arbitrary bipartite splitting, $A,B$, where $A$ ($B$) denotes a subsystem composed out of $N_A$ ($N_B$) qubits respectively. We write the $n$--qubit LMES, $\ket{\Psi}$ as $\ket{\Psi}=U\ket{+}^{\otimes n}$, where $n=N_A+N_B$ and $U$ denotes a phase gate. Since $U$ can be decomposed into pure commuting phase gates, we have $U=U_A U_{AB} U_B$, where $U_A$ ($U_B$) is only acting on the qubits within $A$ ($B$) respectively and $U_{AB}$ is acting non--trivially on both, $A$ and $B$.
Then, the reduced state of system $A$ can be written as $\rho_A=U_A \tr_B(U_{AB} (\proj{+})^{\otimes (n)} U_{AB}^\dagger) U_A^\dagger$. Thus, we will focus in the following on the computation of the reduced state of $\ket{\Psi}=U_{AB}\ket{+}^{\otimes n} \equiv \sum_\i e^{i \alpha_\i}\ket{\i}$. Writing $\ket{\i}=\ket{\i_A}\ket{\i_B}$ we have $\rho_A=\sum_{\i_A,\j_A,\k_B} e^{i(\alpha_{\i_A,\k_B}-\alpha_{\j_A,\k_B})}\ket{\i_A}\bra{\j_A}$. Depending on the correlations of $\alpha_\i$ the reduced state can be computed efficiently, or not. We are going to consider now some simple examples of LMESs for which the reduced state can be computed efficiently. In general, however, this will not be possible, as will be argued below. If $\ket{\Psi}$ is a product state, i.e. $\alpha_\i=\sum_{k}\alpha_{i_k}$, then $\rho_A=\sum_{\i_A,\j_A} e^{i(\alpha_{\i_A}-\alpha_{\j_A})}\ket{\i_A}\bra{\j_A}$ is a pure product state. The second class in the hierarchy of LMESs (see Eq (\ref{Prep})) constitute the so--called weighted graph states (see e.g. \cite{CaBr, HartDu, HeBr}), which include (up to local unitaries) the well--known stabilizer states \cite{Gothesis97}. There, $\alpha_\i$ is a polynomial of degree two in $\i$, i.e. $\alpha_\i=\i^T \Gamma \i$ (up to local unitaries), where the $n\times n$ matrix $\Gamma\equiv\proj{0}\otimes \Gamma_A+\proj{1}\otimes \Gamma_B+(\ket{0}\bra{1}+\ket{1}\bra{0})\otimes \Gamma_{AB} $ is called adjacency matrix. In this case, the reduced state has the form $($up to normalization$)$  \bea \rho_A=\sum_{\vec{s}_A,\vec{s}^\prime_A} e^{i(\vec{s}_A^T \Gamma_A \vec{s}_A- (\vec{s}^\prime_A)^T \Gamma_A \vec{s}^\prime_A)} c_{\vec{s}_A,\vec{s}_A^\prime} \ket{\vec{s}_A}\bra{\vec{s}_A^\prime},\eea

where $c_{\vec{s}_A,\vec{s}_A^\prime}=\prod_{i=1}^{N_B} (1+e^{i(e_i^T \Gamma_{AB} (\vec{s}_A-\vec{s}_A^\prime))})$, with $e_i$ denoting the standard basis. Note that $c_{\vec{s}_A,\vec{s}_A^\prime}$ and therefore $\rho_A$ can be computed efficiently \cite{DuBr05}.

In general, however, it will not be possible to determined the reduced state of an arbitrary LMES efficiently. The reason for this is the following.
Let us consider the $n$--qubit LMES, $\ket{\Psi}=U\ket{+}^{\otimes n}$, where $U$ is a product of pure phase gates of order three, i.e. $\alpha_\i$ is a monomial of degree three in $\i$. Let us now compute the reduced state of qubit $1$. Without loss of generality we write $U=V_1 V_0$, with $V_0$ acting as the identity on the first qubit, and $V_1=\proj{0}_1\otimes \one+ \proj{1}_1\otimes U_1$, where $U_1$ is a product of two--qubit phase gate. Thus, we have $\ket{\Psi}=V_0(\ket{0}_1\ket{+}^{\otimes(n-1)}+\ket{1}_1U_1\ket{+}^{\otimes(n-1)})$.
Writing the weighted graph state, $\ket{\phi}=U_1\ket{+}^{\otimes (n-1)}$ as $\ket{\phi}=\sum_{\vec{s}} e^{i \vec{s}^T \Gamma \vec{s}}\ket{\vec{s}}$ and noting that $\phantom{.}^{(n-1) \otimes}\bra{+}\vec{s}\rangle=\frac{1}{\sqrt{2}^{n-1}}$ $\forall \vec{s}$, we see that we would need to efficiently compute $\sum_{\vec{s}} e^{i \vec{s}^T \Gamma \vec{s}}$ in order to efficiently compute $\rho_1$. It is however very unlikely that this expression can be computed efficiently for an arbitrary $\Gamma$, since it resembles the partition function, which is in general not efficiently computable.

We will use now the discussion above to understand for which states it will be possible to compute the reduced state efficiently. Suppose that $\Gamma$ is block--diagonal, i.e. $\Gamma =\bigoplus \Gamma_i$, where the dimension of each of the block--diagonal matrices, $\Gamma_i$ is independent of $n$. Then, it is easy to see that $\sum_{\vec{s}} e^{i \vec{s}^T \Gamma \vec{s}}$ can be computed efficiently and therefore the reduced state can be computed efficiently. Note that for the state $\ket{\Psi}$ this means that qubit $1$ is independently coupled to several other qubits, but among the set of coupled qubits there is no coupling. That is, the number of overlapping qubits which are coupled to qubit $1$ is small. Similarly, one can see which $m$--partite reduced states of LMES with higher order can be computed efficiently.

Let us now compute the reduced state of certain LMESs, $\ket{\Psi}=U\ket{+}^{\otimes n}$, which will become relevant in the subsequent sections.
For a single qubit, which we choose here without loss of generality to be qubit $1$, and an arbitrary interaction, $U$, which we write as before as $U=V_1 V_0$, with $V_0$ acting as the identity on the first qubit, and $V_1=\proj{0}_1\otimes \one+\proj{1}_1\otimes U_1$, where $U_1=\sum_\i e^{i \alpha_\i }\proj{\i}$ we find $\rho_1=\frac{1}{2}(\one+\ket{0}\bra{1}\phantom{.}^{(n-1) \otimes}\bra{+}U^\dagger_1\ket{+}^{\otimes (n-1)}+h.c.),$ where $\phantom{.}^{(n-1) \otimes}\bra{+}U^\dagger_1\ket{+}^{\otimes (n-1)}=\sum_\i e^{-i\alpha_\i}$. Suppose now that $U_1$ is a product of pure phase gates, $U_l$, each acting on $k_l-1$ different qubits with phases $\phi_l$. Then, the reduced state is given by
\bea \rho_1=\frac{1}{2}(\one+\ket{0}\bra{1}\prod_l \frac{1}{2^{k_l-1}} (2^{k_l-1}-1+e^{-i\phi_l})+h.c.).\eea

As another example we compute the single qubit reduced state of the state $U(\phi)\ket{+}^{\otimes n}$, where $U(\phi)$ is a pure phase gate, i.e. $U(\phi)=\one+(e^{i\phi}-1)\proj{1}^{\otimes n}$. It is easy to show that $\rho_1(\phi)=\tr_{\mbox{all but} 1}(U(\phi)\proj{+}^{n} U(\phi)^\dagger)$ has the following form

\bea  \label{singlered} \rho_1(\phi)=
\frac{1}{2^n} \left(\begin{array}{cc} 2^{n-1}& 2^{n-1}-1+e^{-i\phi} \\ 2^{n-1}-1+e^{i\phi} & 2^{n-1}\end{array}\right).\eea

Similarly the reduced state of several qubits can be determined.

Another way to compute the reduced state is to make use of the notion of projected--pair--entangled state (PEPS) \cite{VeMuCi,VeCipeps}. The idea is to rewrite the quantum state by substituting physical qubits on which several phase gates are acting on by a number of virtual qubits. The transfer from the virtual system to the physical one is done by projectors acting on those virtual qubits. For example, if qubit $i$ is part of $k$ pure phase gates of arbitrary order, we would introduce $k$ virtual qubits, one per phase gate. Then $i_k$ denotes the $k$th virtual qubit of the physical qubit $i$.
The following simple example of two three-qubit phase gates illustrates the idea: $U_{123}U_{345}\ket{+}^{\otimes 5}=P_{3}^{3_{1}3_{2}}U_{123_{1}}U_{3_{2}45}\ket{+}^{\otimes 6}$ with $P_{3}^{3_{1}3_{2}}=\sqrt{2}(\ket{0}_{3}\bra{0 0}_{3_{1}3_{2}}+\ket{1}_{3}\bra{1 1}_{3_{1}3_{2}})$.
The projector, $P$, is independent of the order of the phase gates applied to the physical qubit and independent of the phases. It only depends on the number of pure phase gates which are acting non--trivially on system $3$. In general, the projector associated to a physical qubit $i$ on which $k$ pure phase gates are acting on has the following form $P_{i}^{i_{1},\ldots i_k}=\sqrt{2}^{k-1}(\ket{0}_{i}\bra{00\ldots0}_{i_{1},\ldots i_k}+\ket{1}_{i}\bra{11\ldots1}_{i_{1},\ldots i_k})$ \footnote{Note that using the PEPS--formalism, a LMES can also be written in terms of maximally entangled states and projectors which depend on the phases. For instance, the $3$--qubit LMES, $\ket{\Psi}= U_{123}(\phi_{123})\ket{+}^{\otimes 3}$ can be written as $\ket{\Psi}=p_{3}^{3_{1}3_{2}}(\phi_{123})(\ket{\phi^{+}}_{3_{1}1} \ket{\phi^{+}}_{3_{2}2})$
with  $p_{i}^{i_1 i_2}(\phi)=\sqrt{2} (\ket{0} \bra{++}+\ket{1} \bra{++})-\frac{(1-e^{i\phi})}{\sqrt{2}}\ket{1}\bra{11}$.}.

Let us now use the procedure described above to present the reduced state $\rho_A$ of an arbitrary LMES $\ket{\Psi}_{AB}=U_{AB}\ket{+}^{\otimes (N_A+N_B)}$. Since the phase gates commute, the reduced state can be determined from an interaction pattern without the need of considering any time ordering.
For each qubit in $A$ and $B$, which interacts via several pure phase gates with other qubits, we introduce projectors and virtual qubits as explained above. Note that any pure phase gate is acting on different virtual qubits. That is, there is no virtual qubit on which two or more phase gates are acting on. All information about the coupling is then captured by the projectors. An easy example for $2$--qubit phase gates would be: $U_{12}U_{23}\proj{+}^{3}U^{\dagger}_{12}U^{\dagger}_{23}=
P^{2_1,2_2}_{2}(U_{1 2_1}U_{2_2 3}\proj{+}^{4}U^{\dagger}_{1 2_1}U^{\dagger}_{2_2 3})P^{\dagger 2_1,2_2}_{2}=
P^{2_1,2_2}_{2}(\rho_{1 2_1} \otimes \rho_{2_2 3})P^{\dagger 2_1,2_2}_{2}$,
where $P_{2}^{2_{1},2_2}=\sqrt{2}(\ket{0}_{2}\bra{00}_{2_{1}2_2}+\ket{1}_{2}\bra{11}_{2_{1}2_2})$ and
 $\rho_{ij}=U_{ij}\proj{+}^{2}U^{\dagger}_{ij}$ with $(i,j)\in\{(1 ,2_1), (2_2, 3)\}$. In general, we define for any qubit $b_j$ in $B$, where $m(b_j)$ pure phase gates are acting on, the projector $P^{b_{j_{1}},\ldots b_{j_m(b_j)}}_{b_j}=\sqrt{2}^{m(b_j)-1}(\ket{0}_{b_j}\bra{00\ldots0}_{b_{j_{1}},\ldots b_{j_m(b_j)}}+\ket{1}_{b_j}\bra{11\ldots1}_{b_{j_{1}},\ldots b_{j_m(b_j)}})$ and use a similar definition for any system $s_j$ in $A$. Then the reduced state of system $A$ can be written as
$\rho_ {A}=\tr_{B}(\ket{\Psi}\bra{\Psi})=P_{A} \tr_{B}( P_{B} \otimes_{i} \rho_{i} \otimes_{\{kl\}} \rho_{kl} \ldots \otimes_{\{o, \ldots n\}} \rho_{o,\ldots n} P^{\dagger}_{B})  P^{\dagger}_{A}$, where $P_B=\bigotimes_{j}P^{b_{j_{1}},\ldots b_{j_m(b_j)}}_{b_j}$ and $P_A$ is defined analogously. The operators $\rho$ denote, like in the example above, pure LMESs. For instance, $\rho_{i}$ denotes a pure state which is obtained by applying a $1$-qubit phase gates on the state $\ket{+}$ state, the operators $\rho_{kl}$ result from $2$--qubit phase gates applied to $\ket{++}$ and so forth.We will use this form later to compute the reduced state of special cases.

\section{Decoherence maps}
\label{SecDecMap}

We consider a system composed out of $N_S$ system qubits, which will be denoted by $s_1,\ldots s_{N_S}$, which are interacting with $N_B$ bath qubits, which will be denoted by $b_1,\ldots b_{N_B}$. The qubits interact via some arbitrary phase gate, which will be denoted by $\tilde{U}_{SB}$. The evolution of the system qubits is governed by the completely positive map, $\tilde{{\cal E}}(\rho_S)=\tr_B(\tilde{U}_{SB} \rho_S \otimes \proj{+}^{\otimes N_B} \tilde{U}_{SB}^\dagger)$. In this section we will first of all consider the most general dephasing map and show how it can be represented in terms of the reduced state of some LMES. After that we will represent the dephasing map in terms of Pauli operators. Since we are interested in the decoherence caused by such an interaction we will restrict ourselves in Sec. \ref{SubsecPurelyDephasing} to those interactions where non of the system qubits is directly interacting with another system qubit. As mentioned before, we will call those maps purely dephasing maps. We will show that all of them are separable and derive a simple expression for them. Then we will investigate how the correlations among the bath qubits will affect the evolution of the system qubits.

As before, we decompose the general phase gate, $\tilde{U}_{SB}$, into $\tilde{U}_{SB}=U_B U_{SB} U_S$, where $U_B$ ($U_S$) is acting exclusively on the bath (system) qubits respectively and $U_{SB}$ describes the interaction between the system and bath qubits. Thus, we have $\tilde{{\cal E}}(\rho_S)=U_S \tr_B(U_{SB} \rho_S \otimes \proj{+}^{\otimes N_B} U_{SB}^\dagger)U_S^\dagger$. In order to understand how decoherence can affect the properties of the system state it is therefore sufficient to consider the map ${\cal E}(\rho)=\tr_B(U_{SB} \rho_S \otimes \proj{+}^{\otimes N_B} U_{SB}^\dagger)$. In the following we will focus on this expression.

Note that $U_{SB}$ can always be written as $U_{SB}=\sum_{{\bf k}} \ket{\bf{k}}_B\bra{{\bf k}} \otimes U_{{\bf k}}$, where $U_{{\bf k}}$ denotes a phase gate acting only on the system qubits \footnote{Note that in order to make the corresponding completely positive $($CP$)$ map trace preserving we would simply need to consider $1/2^{N_B} U_{SB}$. In the following we will however not explicitly write this normalization factor.}. Using this decomposition it is easy to see that
\bea \label{map}{\cal E}(\rho_S)=\sum_{{\bf k}}  U_{{\bf k}} \rho_S U_{{\bf k}}^\dagger=\rhosigma \bigodot \rho_S,\eea

where $\rhosigma= \sum_{{\bf k}}  U_{{\bf k}} \proj{+}^{\otimes N_S} U_{{\bf k}}^\dagger$ and $\odot$ denotes the Hadamard product \cite{HornJo}. Note that $\rhosigma={\cal E}(\proj{+}^{\otimes N_S})$ is the $($unnormalized$)$ reduced state of the LMES $U_{SB} \ket{+}^{\otimes n}$, with $n=N_S+N_B$. In order to obtain the second equality in Eq. (\ref{map}) we made use of the fact that for any $n$--qubit state, $\rho$ and any phase gate, $U$, the following holds $U\rho U^\dagger=[U\proj{+}^{\otimes n} U^\dagger] \bigodot \rho$. Maps which can be written as in Eq (\ref{map}) are called diagonal or Hadamard maps. For these maps several interesting properties in the context of optimal output entropy and purity have been recently shown \cite{King}.

It is important to note that the evolution of the system is completely determined by the reduced state $\rhosigma$. Thus, all the correlations which are established between the system and the bath during the evolution are completely captured by the reduced state of the corresponding LMES, $U_{SB} \ket{+}^{\otimes n}$. As mentioned above, the computation of this reduced state gets more and more complicated as the number of overlapping edges, or qubits increases. Physically this corresponds to the situation where more and more bath qubits take part in joint collisions with system qubits.

Another way of representing the dephasing map is to write it in terms of Pauli operators. To do so, we recall the so--called Choi--Jamiolkowski isomorphism between completely positive maps and positive semidefinite operators (states). For a completely positive map, ${\cal E}:{\cal B}({\cal H}_1)\rightarrow {\cal B}({\cal H}_1)$, the corresponding state, $E_{12}$, is acting on the Hilbert space ${\cal H}_1\otimes {\cal H}_1$. It is defined as $E_{12}=({\cal E}_1\otimes \one_2)( P_{12})$, where $P=\proj{\Phi^+}$ with $\ket{\Phi^+}=\sum_{i}^{\mbox{dim}({\cal H}_1)}\ket{ii}$. On the other hand, we have, ${\cal E}(\rho)\propto \tr_{23}(E_{12}\rho_3 P_{23})$. In the case of dephasing maps, the corresponding Choi--Jamiolkowski state, $E$, has the simple form $E_{s_1, \ldots s_{N_S}, a_1, \ldots a_{N_S}} =\E_{s_{1}, \ldots s_{N_S}}\otimes \one_{a_{1}, \ldots a_{N_S}} (\bigotimes_{i=1}^{{N_S}} \ket{\phi^+}_{s_i a_i}\bra{\phi^+})= \bigotimes_{i} \tilde{P}_{s_i a_i} (\rhosigma\otimes \proj{+}^{\otimes N_{S}})\bigotimes_{i} \tilde{P}^{\dagger}_{s_i a_i}$, where $ \tilde{P}_{s_i a_i}=\sqrt{2}(\proj{00}+\proj{11})$ and $a_i$ denote the auxiliary systems. That is, the Choi-Jamiolkowski state $E$ can be rewritten in terms of the reduced state $\rhosigma$ of the LMES, $U_{SB} \ket{+}^{\otimes n}$.

We will now use the operator $E$ to rewrite the map in terms of the Pauli operators. We decompose $E$ in the Bell basis, $E=\sum_{i_1,\ldots ,i_n,j_1,\ldots, j_n} \lambda_{i_1,\ldots ,i_n}^{j_1,\ldots, j_n} \ket{\Psi_{i_1},\ldots,\Psi_{i_n}}\bra{\Psi_{j_1},\ldots, \Psi_{j_n}}$, where $\ket{\Psi_{i_k}}=\sigma_{i_k}\otimes \one \ket{\Phi^+}$, with $i_k\in\{0,\ldots,3\}$ and $\sigma_0=\one$, $\sigma_1=\sigma_{x}$, $\sigma_2=\sigma_{y}$, $\sigma_3=\sigma_{z}$ denote the Pauli operators. In order to present the map in a compact form we use the notation $ \ket{\Psi_\i}=\ket{\Psi_{i_1},\ldots,\Psi_{i_n}}$. Using now that ${\cal E}(\rho)\propto \tr_{23}(E_{12}\rho_3 P_{23})$ we find \bea \label{mapPauli} {\cal E}(\rho)\propto \sum_{\i,\j} \lambda_{\i}^{\j} (\bigotimes_{k=1}^n\sigma_{i_k})\rho (\bigotimes_{l=1}^n\sigma_{j_l}),\eea with $\lambda_{\i}^{\j}=\bra{\Psi_\i}E\ket{\Psi_\j}$. Due to the projector $\tilde{P}$ occurring in the expression of $E$, which projects onto the subspace spanned by $\{\ket{\Psi_0},\ket{\Psi_3}\}$ we find that $\lambda_{\i}^{\j}=0$ unless $i_k,j_l\in \{0,3\}$ for all $1\leq k, j \leq n$.

Let us now consider some simple examples. First of all we consider the simplest case where a single system qubit is coupled to several bath qubits via two--qubit phase gates. This scenario was already treated in \cite{HartDu}. There, the decoherence map for the system qubit, denoted here by  $1$, which interacts with some bath qubits $b_i \in B_1$ is given by
$\E(\rho)=\tr_{B}(\prod_{b_i\in B_1}U_{1b_i} (\rho_1 \otimes \ket{+}_{b_i} \bra{+}) \prod_{b_i} U^{\dagger}_{1b_i})
=\lambda^{0}_{0} \rho + \lambda^{1}_{1} \sz \rho \sz + \lambda^{1}_{0} (\one \rho \sz-\sz \rho \one)$,
with $\lambda^{0}_{0}=(1+r \cos \gamma)/2$, $\lambda^{1}_{1}=(1-r \cos \gamma)/2$ and $\lambda^{1}_{0}=(i r \sin \gamma)/2$ where $r=\prod_{b_i \in B} \cos(\phi_{1b_i}/2)$ and $\gamma=\sum_{b_i \in B}(\phi_{1b_i}/2)$.
The easiest example when considering $3$--qubit interactions describes the situation where a single $3$--qubit phase gate is acting on one system particle and two bath qubits denoted by $b_i$ $b_j$. We find
\bea \E(\rho)&=&\tr_{B}(U_{1 b_i b_j} (\rho \otimes \ket{++}_{b_i b_j} \bra{++}) U^{\dagger}_{1 b_i b_j}) \nonumber \\
&=& \lambda^{0}_{0}\rho-\lambda^{3}_{3} \sz\rho\sz+
\lambda_{0}^{3}(\one \rho \sz-\sz\rho\one) \label{Pauli1} \eea
with $\lambda^{0}_{0}=\frac{1}{8}(7+ \cos(\phi))$, $\lambda^{3}_{3}=\frac{1}{8}(1-\cos(\phi))$ and $\lambda^{3}_{0}=\frac{i}{8}\sin(\phi)$.\\
 Investigating now a sequence of interactions with three--qubit phase gates forces us to consider different cases, namely the cases of independent collisions or collisions with overlapping qubits/edges. We will discuss the maps describing the different coupling scenarios in Sec. \ref{SecDiffInter}.\\
For two system qubits where each of them couples to the bath qubits via one three--qubit phase gate we end up with three different cases.
When the bath qubits are non-overlapping we have $
\E(\rho)=\E_{1} \otimes \E_{2}(\rho)$
with $\E_i$ denoting the single qubit map for a three qubit phase gate (see Eq. (\ref{Pauli1})). If the system qubit $1$ is interacting with two bath qubits, say $b_1$ $b_2$ and system qubit 2 with $b_2$ and $b_3$, i.e. $b_2$ denotes an overlapping qubit we find
$\E(\rho)=\tr_{b_1 b_2 b_3}(U_{1b_1 b_2}U_{2b_2 b_3}(\rho \otimes (\proj{+})^{\otimes 3})U^{\dagger}_{1b_1 b_2}U^{\dagger}_{2b_2 b_3}) =\lambda_{00}^{00}\rho+\lambda_{00}^{30}(\rho\sz\one-\sz\one\rho)+
\lambda_{00}^{03}(\rho\one\sz -\one\sz\rho)+\lambda_{30}^{30}(\sz\one\rho\sz\one)+
\lambda_{03}^{30}(\one\sz\rho\sz\one +\sz\one\rho\one\sz-\sz\sz\rho-\rho\sz\sz)+
\lambda_{03}^{03}(\one\sz\rho\one\sz)+\lambda_{30}^{33}(\sz\one\rho\sz\sz-\sz\sz\rho\sz\one)+
\lambda_{03}^{33}(\one\sz\rho\sz\sz-\sz\sz\rho\one\sz)+\lambda_{33}^{33}(\sz\sz\rho\sz\sz)$
with $\lambda_{00}^{00}=\frac{1}{32}(25+3 \cos(\phi_{2})+\cos(\phi_{1})(3+\cos(\phi_{2})))$,
$\lambda_{00}^{30}=\frac{i}{32}(3+\cos(\phi_{2}))\sin(\phi_{1})$,
$\lambda_{00}^{03}=\frac{i}{32}(3+\cos(\phi_{1}))\sin(\phi_{2})$,
$\lambda_{30}^{30}=\frac{1}{16}(3+\cos(\phi_{2}))\sin^{2}(\phi_{1}/2)$,
$\lambda_{03}^{30}=\frac{1}{32}(\sin(\phi_{1})\sin(\phi_{2}))$,
$\lambda_{03}^{03}=\frac{1}{16}(3+\cos(\phi_{1}))\sin^{2}(\phi_{2}/2)$,
$\lambda_{30}^{33}=\frac{i}{16}(\sin^{2}(\phi_{1}/2) \sin(\phi_{2}))$,
$\lambda_{03}^{33}=\frac{i}{16}(\sin(\phi_{1}) \sin^{2}(\phi_{2}/2))$,
$\lambda_{33}^{33}=\frac{1}{8}(\sin^{2}(\phi_{1}/2)\sin^{2}(\phi_{2}/2))$.
For the case where $b_3$ coincides with $b_1$, i.e. when there is an overlapping edge, the structure of the map is the same as in the overlapping qubit case. The coefficients however change to $\lambda_{00}^{00}=\frac{1}{16}(13+ \cos(\phi_{2})+\cos(\phi_{1})(1+\cos(\phi_{2})))$ and
$\lambda_{00}^{30}=\frac{i}{8}(\cos^{2}(\phi_{2}/2))\sin(\phi_{1})$ and
$\lambda_{00}^{03}=\frac{i}{8}(\cos^{2}(\phi_{1}))\sin(\phi_{2})$
$\lambda_{30}^{30}=\frac{1}{4}\cos^{2}(\phi_{2}/2)\sin^{2}(\phi_{1}/2)$
$\lambda_{03}^{30}=\frac{1}{16}(\sin(\phi_{1})\sin(\phi_{2}))$
$\lambda_{03}^{03}=\frac{1}{4}(\cos^{2}(\phi_{1}/2))\sin^{2}(\phi_{2}/2)$
$\lambda_{30}^{33}=\frac{i}{8}(\sin^{2}(\phi_{1}/2) \sin(\phi_{2}))$
$\lambda_{03}^{33}=\frac{i}{8}(\sin(\phi_{1}) \sin^{2}(\phi_{2}/2))$
$\lambda_{33}^{33}=\frac{1}{4}(\sin^{2}(\phi_{1}/2)\sin^{2}(\phi_{2}/2))$.

In summary, we have shown here three different ways of representing dephasing maps, namely in terms of the Kraus operators (see Eq. (\ref{map})), as a Hadamard product with the reduced state $\rhosigma$ (see Eq. (\ref{map})) and in terms of the Pauli operators (see Eq. (\ref{mapPauli})). In the following we will mainly use Eq. (\ref{map}) to investigate the effects of decoherence.

\subsection{Purely dephasing maps}

\label{SubsecPurelyDephasing}

Let us now focus on those interactions where each system qubit is only interacting with bath qubits, but not with another system qubit. We will call these maps purely dephasing maps, because, as we will show, they are all separable, i.e. they are of the from ${\cal E}(\rho)=\sum_i p_i A_i\otimes B_i \otimes \cdots \otimes N_i \rho A^\dagger_i\otimes B^\dagger_i \otimes \cdots \otimes N^\dagger_i$. Note that those maps cannot generate entanglement and can be implemented (at least probabilistically) by local operations and classical communication. The reason, why we are especially interested in those maps is because we are interested in the decoherence effect of multipartite phase gates and not in generating entanglement among the system qubits. Apart from that, considering those maps allows us to make a fair comparison  to the decoherence caused by two-qubit phase gates, which always leads to separable maps.

In this section we will first show that any purely dephasing map is separable. Then we will restrict ourselves to those situations where $m$ bath qubits are interacting with all qubits in the system. The system qubits are also interacting with some additional, but not overlapping qubits. We will present a simple formula for those maps, which will allow us to study how bath correlations affect the evolution of the system. This kind of investigations will further allow us to determine the behavior of the system in the case of Markovian and non--Markovian interactions.

As before we consider the map ${\cal E}(\rho)=\tr_B(U_{SB} \rho_S \otimes (\proj{+})^{\otimes N_B} U_{SB}^\dagger)$. In order to deal with the various interactions we are going to introduce the following notation. Let $I=\{i_1,\ldots, i_k\}$, denote a set of $k$ indices for some $k\leq N_B$ and $B_I$ denote those $k$ bath qubits, i.e. $B_I\equiv \{b_{i_1},\ldots, b_{i_k}\}$. Then, $U_{B_I,s_l}$ denotes a pure phase gate acting on the bath qubits,
$B_I$ and on the system qubit $s_l$. Furthermore, ${\cal I}$ denotes the set of all sets of indices of bath qubits which are interacting with some system qubit, i.e. ${\cal I}=\{I \in \{1,\ldots, N_B\}^{\times k}, \mbox{for some } k\leq N_B \mbox{ s.t. } \exists l \mbox{ s.t. }  U_{B_I,s_l}\neq \one\}$. That is, for each $I\in {\cal I}$ there exist $|I|$ bath qubits $B_I$ and the single system qubit, $s_l$ which interact with each other via the non--trivial pure phase gate of order $|I|+1$, $U_{B_I,s_l}$. For each system qubit, $l$, ${\cal I}_l\subset {\cal I}$ denotes the set of indices, $I=\{i_1,\ldots, i_k\}$, for some $k\leq N_B$ such that the bath qubits $B_I=\{b_{i_1},\ldots, b_{i_k}\}$ interact with the system qubit $l$. That is, ${\cal I}_l=\{I\in \{1,\ldots, N_B\}^{\times k}, \mbox{for some } k\leq N_B \mbox{ s.t. } U_{B_I,s_l}\neq \one\}$. Moreover, for each set of indices, $I$, we introduce the function $d_I(\k)=\prod _{i\in I} k_i$. The pure phase gates $U_{B_I,s_l}$ can be written as $U_{B_I,s_l}=\one + \proj{1,\ldots,1}_{B_I}\otimes (U_{I,s_l}-\one)$ where $U_{I,s_l}$ is a single qubit phase gate acting on the system qubit $s_l$ \footnote{Note that any phase gate can be decomposed in this form.}. Thus, we have $U_{B_I,s_l}=\sum_\k \proj{\k}_{B_I} \otimes U_{I,s_l}^{d_I(\k)}\otimes \one$, where $U_{I,s_l}^{d_I(\k)}=\one$ if $d_I(\k)=0$ and $U_{I,s_l}^{d_I(\k)}=U_{I,s_l}$ if $d_I(\k)=1$. Since the global unitary, $U_{SB}$ can be written as a product of the unitaries  $U_{B_I,s_l}$, i.e. $U_{SB}=\prod_l  \prod_{I\in {\cal I}_l}U_{B_I,s_l}$ we find

\bea U_{SB}=\sum_\k \proj{\k}_B\otimes U_S^\k,\eea where \bea \label{sepMap}U_S^\k=\bigotimes_{l} \prod_{I\in {\cal I}_l} U_{I,s_l}^{d_I(\k)}\equiv U_1^\k\otimes U_2^\k \otimes \cdots \otimes U_{N_S}^\k,\eea with $U_l^{\k}=\prod_{I\in {\cal I}_l} U_{I,s_l}^{d_I(\k)}$. The corresponding completely positive $($CP$)$ map is given by \bea \label{sepMap1} {\cal E}(\rho)=\sum_\k U_S^\k \rho (U_S^\k)^\dagger=\rhosigma \bigodot \rho,\eea where all $U_S^\k$ given in Eq. (\ref{sepMap})  are local unitary phase gates and $\rhosigma=\sum_\k U_S^\k (\proj{+})^{\otimes {N_S}} (U_S^\k)^\dagger$, is a convex combination of product LMESs and is, in particular, separable (see also Eq (\ref{redpeps}) below). It should be noted here that since the bath qubits can be in an entangled state, there is no reason, in general, why the map should be separable. However, in the case investigated here, all the interactions commute, which implies that instead of considering an entangled initial state of the bath qubits, we could equally well consider the product state, $\ket{+}^{\otimes N_B}$ as initial state. This fact alone is however, not enough to conclude that the maps are always separable, since the baths qubits couple to several system qubits.

So far we have seen that, whenever each system qubit interacts only with bath qubits any decoherence map, $\E$ can be expressed as a convex combination of local decoherence maps, i.e.  $\E(\rho)=\sum_{i} $$p_{i} \E_{i}(\rho)$. Here, $\E_{i}$ denotes a local unitary map on the system qubits. The structure of ${\cal E}_i$ is determined by the number of qubits the phase gates are acting on and by the number of overlapping qubits/edges. As a next step we will investigate the influence of the coupling between the different baths (for each system qubit) on the evolution of the system qubits. The more bath qubits interact with all system qubits the stronger the correlation between the baths (for each system qubit) is. We will derive for any of these correlations a simple expression for the corresponding map, which will allow us then to analyze this effect quantitatively.

We denote by $\E_{m}(\rho)$ the state of the system qubits after each of them interacted via a phase gate (not necessarily pure) of arbitrary order, $n_l+1$, with the bath qubits $\{b_{i_1},\ldots, b_{i_m}\}$ and some other $n_l-m$ bath qubits, which do not interact with any other system qubit. That is, $\E_{m}$ denotes the map which results from an interaction $U_{SB}=\prod_l \prod_{I\in {\cal I}_l}U_{B_I,s_l}$, where ${\cal I}$ is such that the following condition is fulfilled. There exists some fixed set of $m$ indices, $\{i_1,\ldots, i_m\}$, such that
 \bi
 \item[a)] $\forall I\in{\cal I}$, $\{i_1,\ldots, i_m\}\subseteq I$ and
 \item[b)]$\forall $ $I\in {\cal I}_l$, $ I^\prime \in {\cal I}_{l^\prime}$ with $l\neq l^\prime$ it holds that $I\cap I^\prime=\{i_1,\ldots, i_m\}$.
 \ei
For instance, $\E_{0}(\rho)$ will describe the system qubits after all of them interacted with independent bath qubits. $\E_{1}(\rho)$ describes a situation where all the gates acting on the system qubits are coupled to one bath qubit, etc.. Considering these kind of maps allows us to study how the effect of decoherence changes with the degree of dependency of the environments of the system qubits. For instance, how does the decoherence affect the behavior of the system if all system qubits interact with the same bath, compared to the one, where each system interacts with its own bath. In order to study all those cases we are going to derive now a simple expression for the corresponding maps. To this end we introduce the following notation. ${\cal J}_l=\{J \in \{1,\ldots ,N_B\}^k \mbox{ s.t. } \exists I\in {\cal I}_l s.t. J\cup \{1,\ldots, m\}=I\}$, i.e. each $J\in {\cal J}_l$ denotes the bath qubits, $B_J$ which interact with the bath qubits $\{b_1,\ldots b_m\}$ and the $l$--th system qubit via a pure phase gate. ${\cal J}$ denotes the union of all sets ${\cal J}_l$. Let $J=\{j_1,\ldots j_{|J|}\}$ be some set of indices. For some vector $\k=(k_1,\ldots, k_{N_B})$ we denote by $\k_J$ the $|J|$--dimensional vector with entries $\k_j$ for $j\in J$, i.e. $\k_J=(k_{j_1},\ldots k_{j_{|J|}})$. Moreover $\k_{J_l}$ will denote the set of entries of $\k$, $\{k_{j_1},\ldots ,k_{j_r}\}$, where for each $j_t$ with $t\in \{1,r\}$ there exists some $J\in {\cal J}_l$ such that $j_t\in J$. Using this notation we can now state the following theorem.

\begin{theorem} Let $U_{SB}=\prod_l \prod_{I\in {\cal I}_l}U_{B_I,s_l}$, where $B_I$ denotes $|I|$ bath qubits and $s_l$ denotes a single system qubit. Moreover, let ${\cal I}$ be such that the conditions (a) and (b) are fulfilled for some set of indices, $\{i_1,\ldots, i_m\}$. Then the corresponding CP map, ${\cal E}=\tr_B[U_{SB} (\rho \otimes \proj{+}^{\otimes N_B} ) U_{SB}^\dagger]$ is given by
\bea {\cal E}(\rho)=\frac{1}{2^{N_B}}((2^{N_B}-2^{N_B-m}) \rho+\tilde{{\cal E}}_1\otimes \ldots \otimes \tilde{{\cal E}}_{N_S}(\rho)), \label{locmaps}\eea

where $\tilde{{\cal E}}_l(\sigma)=\sum_{\k_{J_l}} \tilde{U}_{s_l}^{\k_{J_l}} \sigma (\tilde{U}_{s_l}^{\k_{J_l}})^\dagger.$
Here, $\tilde{U}_{s_l}^{\k_{J_l}}=\prod_{J\in {\cal J}_l} U_{s_l,J}^{d_{J}(\k)}$ is a local unitary operation. \end{theorem}

Note that this implies that the number of bath qubits which are interacting with all system qubits, $m$, only changes the factor in front of $\rho$ and the weight with which $\tilde{U}_{s_l}^\k$ is applied but leaves the rest unchanged.

\begin{proof}
We assume without loss of generality that the set of bath qubits with which any system qubit is interacting, $\{b_{i_1},\ldots, b_{i_m}\}$, is $\{b_{1},\ldots, b_{m}\}$ (condition (a)). Then, due to the fact that $\{1,\ldots, m\}\subseteq I$ $\forall I\in {\cal I}$ we have that $d_I(\k)=\prod_{i=1}^m k_i \prod_{i\in I\diagdown \{1,\ldots, m\}} k_i\equiv c(\k) \prod_{i\in I\diagdown \{1,\ldots, m\}} k_i$, where $c(\k)= \prod_{i=1}^m k_i$ is independent of $I$. Thus, we find for the local unitary transformations, $U_l^\k$ in Eq (\ref{sepMap}), $U_l^\k=(\tilde{U}_l^\k)^{c(\k)}$, where $\tilde{U}_l^\k=\prod_{I\in {\cal I}_l} U_{I,s_l}^{d_{I\diagdown \{1,\ldots, m\}}(\k)}$. Using this expression in Eq (\ref{sepMap1}) and the fact that $\sum_{\k:c(\k)=0} 1= 2^{N_B}-2^{N_B-m}$ we find ${\cal E}(\rho)= \frac{1}{2^{N_B}}[(2^{N_B}-2^{N_B-m}) \rho+\tilde{{\cal E}}(\rho)]$, where $\tilde{{\cal E}}(\rho)=\sum_{\k=(k_{m+1},\ldots,k_{N_B})} \bigotimes_l \tilde{U}_l^\k\rho \bigotimes_l (\tilde{U}_l^\k)^\dagger$. As mentioned before, ${\cal J}$ denotes the set of indices of baths qubits which are interacting with the system qubits and the bath qubits $b_1,\ldots,b_m$. Hence, for each pair $J\in {\cal J}_l$ and $J^\prime \in {\cal J}_{l^\prime}$ with $l\neq l^\prime$ we have $J\bigcap J^\prime= \emptyset$. Thus, the sum in the expression of $\tilde{{\cal E}}(\rho)$ can be decomposed as a sum over all bath qubits with are interacting with the first system qubit, ${\cal J}_1$, the second, ${\cal J}_2$, and so on. Since non of the indices will occur twice, we have $\tilde{{\cal E}}(\rho)=\tilde{{\cal E}}_1\otimes \ldots \otimes \tilde{{\cal E}}_{N_S}(\rho)),$ with $\tilde{{\cal E}}_l$ as defined above.

\end{proof}

It should be noted here that this result is independent of the coupling between the bath qubits and a single system qubit. The influence of their couplings and orders is reflected in the details of the local maps $\tilde{{\cal E}}_l$, but does not change the general form of the map.

In summary, we have seen that whenever some ($m$) of the bath qubits are interacting with all system qubits via some arbitrary pure phase gates and if no other overlapping bath qubit exists, then the map corresponds to a convex combination of the identity  (with a weight that depends on $m$) and a local map. The local map corresponds to the map one would get by conditioning on the fact that all qubits $b_1,\ldots b_m$ are in the state $\ket{1}$.
That is, $\tilde{{\cal E}}_{l}(\sigma)=\tr_{B}(U_{B_{I\diagdown \{1,\ldots,m\}},s_l} [\sigma\otimes (\proj{+})^{B_{I\diagdown \{1,\ldots,m\}}}]   U_{B_{I\diagdown \{1,\ldots,m\}},s_l}^\dagger)$, where $U_{B_{I\diagdown \{1,\ldots,m\}},s_l}=_{b_1,\ldots b_m}\bra{1,\ldots,1}\prod_{I\in {\cal I}_l}U_{B_I,s_l}\ket{1,\ldots 1}_{b_1,\ldots b_m}$.

Let us now consider some examples. We will assume that any system qubit, $l$, is interacting with the bath qubits only via a single pure phase gate of order $n_l+1$ and phase $\phi_l$. Suppose further that the conditions (a) and (b) are fulfilled for some set of $m$ indices, $\{i_1,\ldots, i_m\}$. Let ${\cal E}_m^{n_1,\ldots, n_{N_S}}$ denote the corresponding map. It is straightforward to see that the maps $\tilde{{\cal E}}_l$ in Eq (\ref{locmaps})
are then given by $\tilde{{\cal E}}_l(\sigma)=2^{n_l-m}[(1-2^{m-n_l})\sigma+2^{m-n_l}U_l(\phi_l)\sigma U_l(\phi_l)^\dagger]$. Here and in the following $U_l(\phi_l)$ denotes the single qubit phase gate with phase $\phi_l$. Using then that $\prod_l 2^{n_l-m}=2^{N_B-m}$ we find
\bea {\cal E}_m^{n_1,\ldots, n_{N_S}}(\rho)=\frac{1}{2^{m}}[(2^{m}-1) \rho+\bigotimes_{l=1}^{N_S}{\cal E}_l(\rho)], \label{locmapszwei}\eea
where $\E_{l}(\sigma)=(1-2^{m-n_l})\sigma+2^{m-n_l}U_l(\phi_l)\sigma U_l(\phi_l)^\dagger$. For instance, if all system qubits interact with the same $m$ bath qubits (extreme non--Markovian case, see Sec. \ref{SecMarkovian}) we obtain ${\cal E}_m^{m,\ldots, m}(\rho)=\frac{1}{2^{N_B}}[(2^{N_B}-2^{N_B-m}) \rho+2^{N_B-m}({\cal E}_1\otimes \ldots \otimes {\cal E}_{N_S})(\rho)],$ where $\E_{l}(\sigma)=U_l(\phi_l)\sigma U_l(\phi_l)^\dagger)$. We are going to show in Sec. \ref{SecDiffInter} how the effect of decoherence depends on the number of bath qubits which are interacting with all system qubits. That is, we will analyze there how the degree of correlations between the environment of the various system qubits is affecting the evolution of the system.

Note that any map which results from a pure phase gate of arbitrary order can be reinterpreted as a mixture of the identity map and a map resulting from a two--qubit phase gate acting on the system and the bath qubit, which is prepared in the state $\ket{\Psi}=\sqrt{\alpha} \ket{0}+\sqrt{1-\alpha^2} \ket{1}$, for some appropriately chosen $\alpha \geq 0$. That is, $\E_{l}(\sigma)=(1-2^{m-n_l})\sigma+2^{m-n_l}U_l(\phi_l)\sigma U_l(\phi_l)^\dagger=\tr_B[U_{SB} (\proj{\Psi}\otimes \sigma) U_{SB}^\dagger]$, where $U_{SB}$ denotes the two--qubit pure phase gate with phase $\phi_l$ and $\alpha=1-2^{m-n_l}$.

Let us now also express the maps studied above in terms of the Hadamard product, ${\cal E}(\rho)=\rhosigma \bigodot \rho$. First of all, we present the state $\rhosigma$, for the situation when there is no interaction between the system qubits (see Eq (\ref{sepMap1})).
We start with the LMES $\ket{\psi}=U_{SB}\ket{+}^{N_S+N_B}$.
As the next step we use the PEPS picture and introduce virtual qubits $b_{i_{1}}, \ldots, b_{i_{k}}$ for every bath qubit $b_i$ on which $k$ pure phase gates are acting on. For the qubits which are non-overlapping, which means just one pure phase gate is acting on them, the projector $P^{b_{i_{1}}, \ldots, b_{i_{k}}}_{b_i}$ will simply be the identity and for the overlapping bath qubits it will have the form $P^{b_{i_{1}}, \ldots, b_{i_{k}}}_{b_i}=\sqrt{2}^{k-1}(\ket{0}_{b_i}\bra{00\ldots0}_{b_{i_{1}},\ldots b_{i_k}}+\ket{1}_{b_i}\bra{11\ldots1}_{b_{i_{1}},\ldots b_{i_k}})$.
For $I\in {\cal I}_l$, $B^{\textbf{v}}_{I}$ denotes the set of virtual bath qubits which interact with system qubit $s_l$ via a phase gate $U_{s_l B^{\textbf{v}}_{I}}$.
The LMES in the PEPS picture can be written as
$\ket{\psi}=\bigotimes P^{b_{i_{1}}, \ldots, b_{i_{k}}}_{b_i} \bigotimes_{l} \bigotimes_{I \in {\cal I}_l} \ket{\psi_{s_l,B^{\textbf{v}}_{I}}}$
with $\ket{\psi_{s_l,B^{\textbf{v}}_{I}}}=U_{s_l B^{\textbf{v}}_{I}}\ket{+}^{\vert I \vert +1}$.
As a first step we trace over all qubits which are non-overlapping. We denote by $F=\{i : P^{b_{i_{1}}, \ldots, b_{i_{k}}}_{b_i}=\one_{b_i}\}$ the set of non-overlapping qubits. Then we have $\rhosigma=tr_{B}(\proj{\psi})=\tr_{B\backslash F} [\bigotimes_{i \not \in F} P^{b_{i_{1}}, \ldots, b_{i_{k}}}_{b_i}\otimes_{l} \rho_{s_l,B^{\textbf{v}}_{l}}\bigotimes_{i \not \in F} P^{\dagger b_{i_{1}}, \ldots, b_{i_{k}}}_{b_i}]$ with $\rho_{s_l,B^{\textbf{v}}_{l}}=\tr_{i \in F}(\otimes_{I \in I_l} \ket{\psi_{s_l, B^{\textbf{v}}_{I}}}\bra{\psi_{s_l, B^{\textbf{v}}_{I}}})$.
Performing the trace over all coupled qubits leads to a summation over all operators $\bigotimes_{l} \rho^{\textbf{k}}_{s_l}$ where $\rho^{\textbf{k}}_{s_l}=\bra{\textbf{k}}\rho_{s_l,B^{\textbf{v}}_{l}} \ket{\textbf{k}}$, with $\ket{\textbf{k}}$ denoting the computational basis of the overlapping bath qubits. We end up with \bea \rhosigma=\sum_{\textbf{k}}\bigotimes_{l} \rho^{\textbf{k}}_{s_l} \label{redpeps}\eea
Examples of the reduced state in this form are given in Sec. \ref{SecComparison}. For the special case where all system qubits couple to the same $m$ bath qubits via one pure phase gate $($see Eq. (\ref{locmapszwei})$)$ we find $\rhosigma=\frac{1}{2^m}[(2^m-1)(\proj{+})^{\otimes N_S}+\bigotimes^{N_S}_{l=1} ((1-2^{m-n_l})\proj{+}+2^{m-n_l} U_l \proj{+} U_l^{\dagger})]$\footnote{Note that this corresponds to a map, which is a convex combination of the identity map and the one which results from 2--qubit phase gates, as it should be.}.
As mentioned before, for any $n$--qubit phase gate, $U$ it holds that $U\proj{+}^{\otimes n} U^\dagger \bigodot \rho =U\rho U^\dagger$. Thus, writing $\rhosigma$ as $\rhosigma=\sum_\k U_\k \proj{+}^{\otimes n} U_\k^\dagger$ as we did here, leads directly to the map ${\cal E}(\rho)= \sum_\k U_\k \rho U_\k^\dagger$. Hence, ${\cal E}$ can be directly read off given $\rhosigma$ and visa versa. Depending on which property of the output state one is interested in, one, or the other way of looking at the map is better suited.

\section{Comparison of different interactions}
\label{SecComparison}
In this section we will use the result derived above in order to compare different decoherence models with each other. Due to the large variety of different interactions there is of course no way to make a complete comparison. However, we will focus here on certain relevant aspects of the dependency of the evolution of the system qubits on various couplings. In Sec \ref{SecMarkovian} and Sec \ref{SecDiffInter} we investigate how the decoherence depends on the degree of bath--correlations. In Sec \ref{SecMarkovian}, we will compare the case where there is no dependency, that is, each qubit interacts with an independent bath, and all pure phase gates are applied to fresh bath qubits and the system qubits (Markovian case) to the other extreme case, where the system qubits are always interacting with the same bath qubits (extreme non--Markovian case). We will determine the coherence time for an individual system qubit and compare the two extreme cases for different orders of the applied phase gates. In Sec \ref{SecDiffInter} we will first determine the evolution of a single system qubit and will investigate how the decoherence depends on the degree of correlation of the baths. In contrast to Sec \ref{SecMarkovian} we keep the order of the phase gate fixed and vary the number of qubits which are interacting more than once with the system qubit. Then we will consider $2$--qubit states and certain $n$--qubit states and will show how the evolution of the entanglement shared by the system qubits depends on the coupling within the bath.
Due to Eq. (\ref{locmaps}) one might expect that the system gets less disturbed the larger $m$ is. However the examples investigated here will show that this is not the case. In fact we will show that for certain states the entanglement is more robust against decoherence induced by individual baths $(m=0)$ than against the one where $m$ is very large. In Sec \ref{SecOrder} we will discuss the effect of the order of the applied phase gates on the evolution of the system.

\subsection{Markovian scenario compared to (extreme) non--Markovian scenario}
\label{SecMarkovian}

In the following we will call an interaction, $U_{SB}$, Markovian, if non of the qubits is interacting more than once with any other qubit via a pure phase gate, i.e. $\forall I,I^\prime \in {\cal I}$, $I\cap I^\prime={\O}$. We will compare the decoherence induced by a Markovian process to the one induced by a non--Markovian process, where some bath qubits interact more than once with some system qubits. This comparison will be performed for several different orders of the applied phase gates.

In the Markovian case we have that each system qubit interacts independently from the other system qubits with its own bath. Thus, we have that ${\cal E}(\rho)={\cal E}_1\otimes \ldots \otimes {\cal E}_{N_S}(\rho)$. In order to gain some insight into how this coupling affects the evolution of the system we study how the local maps, ${\cal E}_i$ acts on a single qubit. The single qubit is, at each step, interacting with $n_1$ different bath qubits. We consider here the case where the system qubit is interacting $k$ times via a $(n_1+1)$--qubit phase gate with its environment. The phases of the phase gates are denoted by $\phi_i$, with $i=1,\ldots, k$. The evolution is governed by ${\cal E}(\rho)={\cal E}_k\circ \ldots \circ {\cal E}_1(\rho)$, where ${\cal E}_i(\rho)=\alpha_i \rho+\beta_i U_i(\phi_i)\rho U_i(\phi_i)^\dagger$, with $U_i$ a single qubit phase gate with phase $\phi_i$ and $\alpha_i=1-2^{-n_1}, \beta_i=2^{-n_1}$.
The relevant parameter to be considered here is the off--diagonal element of $\rho$, $\rho_{01}=\bra{0}\rho\ket{1}$, which allows us to determine the coherence time of the system \footnote{As explained in the introduction, the phase gates correspond to a evolution due to the Hamiltonian of the form $H_{SB}=\sum_{\bf i} \alpha_{\bf i} \sigma^{\bf i}_z$. The time dependence of the unitary $U_{SB}=e^{-iH_{SB} t}$ is hidden in the phases of the phase gates. This is why we speak of coherence times, even though we deal with time independent gates.}. Due to Eq. (\ref{map}) it is easy to see that after the evolution, the off-- diagonal element of the output state will be $\rho_{01}\rhosigma^{01}$, where $\rhosigma^{01}=\bra{0}\rhosigma\ket{1}$. For the sake of simplicity we chose now $\phi_i=\phi$ $\forall i \in \{1,\ldots, k\}$. We obtain $\rhosigma={\cal E}(\proj{+})={\cal E}^k_1(\proj{+})=\bigodot_{l=1}^k \rhosigma^1$, where $\rhosigma^1$ corresponds to ${\cal E}_1$ and is given in Eq. (\ref{singlered}) with $n_1=n-1$. Thus, the absolute value of $\rho_{01}$ will be multiplied by the factor
$ \vert  \rhosigma^{01}\vert=\left[\frac{2^{n_{1}}-1}{2^{n_1}}\sqrt{1+\frac{2}{2^{n_1}-1} \cos( \phi)+\frac{1}{(2^{n_1}-1)^{2}}}\right]^{k}$.

Let us now investigate how this factor and therefore the coherence time depends on the order of the applied phase gates. The case where $n_1=1$, i.e. the interaction is described by $2$--qubit phase gates, has been studied in \cite{HartDu}. There, $\vert \rhosigma^{01} \vert=[\cos(\phi/2)]^k$, which can be approximated by $e^{- k \phi^2 /8}$ for small values of $\phi^{2} k$ has been obtained. In case the system qubit interacts via three--qubit phase gates with its bath we have $\vert \rhosigma^{01} \vert=[\frac{1}{4} \sqrt{10+6 \cos(\phi)}]^k \approx e^{-\frac{3}{32}k\phi^2}$. For four--qubit phase gates we get $\vert \rhosigma^{01} \vert=[\frac{1}{8} \sqrt{50+14 \cos(\phi)}]^k \approx e^{-7 k \phi^2/128}$. The coherence in the Markovian process for a single qubit falls therefore off exponentially with respect to the number of occurred collisions, $k$. The coherence time increases as the order of the phase gates get larger. In particular the ratio of the coherence times between the $3$--qubit phase gate case and the $2$--qubit phase gate case is $\frac{32/3}{8} \approx 1,3$ and between the $4$--qubit phase gate case and the $2$--qubit phase gate case is $\frac{128/7}{8} \approx 2,3$.

When the gates are coupled, i.e. the bath qubits interact more than once with the system qubit, we end up with a non-Markovian process, since the processes are not independent of each other. In Sec \ref{SecDiffInter} we will investigate how the evolution depends on the degree of correlation in the bath for a fixed order of applied phase gates. Here, we will consider the extreme case, where the system qubit is interacting always with the same $n_1$ bath qubits. It is easy to see that in this case the evolution is governed by the completely positive map ${\cal E}(\rho)=\rho_{1}(\phi_1+\ldots + \phi_n)\bigodot \rho$, where $\rho_1 (\phi)$ is given in Eq. (\ref{singlered}). There, the coherence term, $\vert \rho_{01} \vert$, gets multiplied by the factor $\vert \rhosigma^{01} \vert=\frac{2^{n_1}-1}{2^{n_1}}\sqrt{1+\frac{2}{2^{n_1}-1} \cos(\sum^{k}_{i=1} \phi_i)+\frac{1}{(2^{n_1}-1)^{2}}}$.
Like in the Markovian case we investigate how the coherence time depends on the order of the applied phase gate. Again, in order to make a simple comparison we choose $\phi_i=\phi$ $\forall i$. The 2-qubit phase gate case was already discussed in \cite{HartDu} where $\vert \rhosigma^{01} \vert= \cos(k \phi/2)\approx e^{- k^2 \phi^{2}/8}$ has been found. The approximation is valid for small values of $k \phi$. Since in the non-Markovian process the sum of all phases appears in a cosine we get an oscillating behavior for large values of $k$. For the three-- and four--qubit interactions we find $\vert \rhosigma^{01} \vert=[\frac{1}{4} \sqrt{10+6 \cos(k \phi)}] \approx e^{- 3 k^2 \phi^{2}/32}$ and $\vert \rhosigma^{01} \vert=[\frac{1}{8} \sqrt{50+14 \cos(k \phi)}]\approx e^{- 7 k^2 \phi^{2}/128}$, respectively. The coherence time is longer if we go to collisions with phase gates of higher order. This is similar to the Markovian case. Within the range of the approximations the type of decay of the coherence in the non-Markovian case is Gaussian whereas the decay in the Markovian case is exponential with respect to the number of collisions $k$. Note that the coefficients in the exponent coincide.

In summary, we see that for both cases, the Markovian and the extremely non--Markovian case, the coherence time increases with the order of the applied phase gates. This can be understood by noting that the higher the order is the more qubits get entangled to the system qubit during the interaction. Since this entanglement is generated by pure phase gates, the entanglement between a single qubit and the rest gets smaller. More precisely, if the single system qubit is interacting with its bath via a pure phase gate of order $n_1+1$, we have ${\cal E}_{n_l}(\rho)=\rho_1$, where $\rho_1$ is given in Eq. (\ref{singlered}). It is easy to see that if $n_1=1$ and $\phi=\pi$, the system qubit is maximally entangled to its bath. However, for $(n_1+1)$--qubit phase gates with $n_1>1$, this is no longer possible since the off--diagonal term of ${\cal E}(\rho)$, $2^{-(n_1+1)}(-1+2^{n_1}+e^{-i\phi})$, is non--vanishing for any value of $\phi$. As mentioned already, the only difference between the two extreme cases investigated here is the dependency on the number of collisions (in the region, where the approximations are valid). Whereas the coherence term in the Markovian case decays exponentially with the number of collisions it does so with a Gaussian function in the extreme non--Markovian case.

\subsection{Comparing different bath dependencies}
\label{SecDiffInter}

In this section we discuss how the induced decoherence depends on how strongly the baths for the individual system qubits are coupled with each other. To this end we keep the order, $n_1+1$, of the applied pure phase gates fixed and the same for each system qubit. That is, each system qubit is interacting with $n_1$ bath qubits. Out of these $n_1$ baths qubits $m$ of them interact with all system qubits. Using the notation from before, we have $\forall I,I^\prime \in {\cal I}$ $I\cap I^\prime=\{b_{i_1},\ldots ,b_{i_m}\}$ for some fixed set of $m$ bath qubits, $\{b_{i_1},\ldots ,b_{i_m}\}$. The aim of this subsection is to investigate how the evolution of the system changes as a function of $m$, the degree of bath--correlation. We will first work out in detail how the decoherence affects the evolution of a single and two system qubits. Then we will consider the general case of $N_S$ system qubits and investigate the influence of the degree of baths--correlation on the evolution. That is, we will use Eq. (\ref{locmaps}) and Eq. (\ref{locmapszwei}) to determine the evolution of the system for the various scenarios. For a single system qubits we consider the situation where the system is interacting with the bath via several phase gates. The number of overlapping qubits, $m$, will denote in this case the number of qubits which are coupled to the system qubit in each phase gate. Physically this means that those $m$ bath qubits take part in any collision with the system qubit. We will investigate here how the decoherence of the system qubit depends on this coupling. For more than a single system qubit we investigate the scenario where each system qubit interacts with its environment via a single pure phase gate. There, $m$ denotes, as explained above, the number of bath qubits which interact with all system qubits. We will also investigate the evolution of entanglement shared between the system qubits.

As mentioned before, depending on which property of the evolved state one is interested in, the presentation of the decoherence map in terms of the Kraus operators, or in the form ${\cal E}(\rho)=\rhosigma \odot \rho$ is better suited. This is why we will give both representations in the examples investigated here. Whenever we talk about the corresponding reduced state in this context, we refer to $\rhosigma$, which denotes the reduced state of the LMES corresponding to the evolution (see Sec. \ref{SecDecMap}). The phase gates, acting on $n$ qubits will be denoted by $U^{n} (\phi_i)=\one-(1-e^{i\phi_i})\proj{1}^{\otimes n}$ and $U_i=U^{1}(\phi_i)$ will denote the single qubit phase gates with phase $\phi_i$.

\subsubsection{A single system qubit}
\label{singqu}
We consider a single system qubit which is interacting with its environment via $k$ different $(n_1+1)$--qubit phase gate.
We denote by $m$ the number of bath qubits, $\{b_1,\ldots ,b_m\}$, which take part in all interactions with the system qubit. That is, $\forall I, I^\prime \in {\cal I}$ with $I\neq  I^\prime$ it holds that $I\cap I^\prime=\{b_1,\ldots ,b_m\}$. Then we have
$\E(\rho)=2^{-m}[(2^m-1)\rho+\circ_{i=1}^k  \E_i (\rho)]$,
where $\circ$ denotes the composition and $\E_i(\rho)=\frac{1}{2^{n_1-m}}((2^{n_1-m}-1)\rho+U_i \rho U^{\dagger}_i).$
Let us now explicitly determine the evolution as a function of $m$. In order to see the effect of the coupling within the bath it suffices to consider only two different phase gates. We start with the Markovian case, where the phase gates are acting on independent bath qubits. The corresponding decoherence map is
$\E(\rho)=\rhosigma \odot \rho=\E_2 \circ \E_1(\rho)$, where
$\E_{j}(\rho)=\lambda_{00} \rho +\lambda_{01} U_j \rho U^{\dagger}_j$ with $\lambda_{00}=1-2^{-n_1}$, $\lambda_{01}=2^{-n_1}$.
Note that the reduced state, $\rhosigma$, is given by $\rhosigma=\rho_1 \odot \rho_2$ where $\rho_i$ is the single qubit reduced state of $U^{n_1+1} (\phi_i) \ket{+}^{\otimes n_1+1}$ (see Sec \ref{redStates}). \\
Next, we investigate the scenario in which one  bath qubit is overlapping. The map can be written as
 $\E(\rho)=\rhosigma \odot \rho=\frac{1}{2}[ \rho+\E_2 \circ \E_1(\rho)]$, where $\E_{j}(\rho)=\lambda_{10}\rho + \lambda_{11} U_j \rho U^{\dagger}_j$ with $\lambda_{10}=1-2^{1-n_1}$ and $\lambda_{11}=2^{1-n_1}$.
The reduced state for this case is given by
$\rhosigma=2\sum_{k=0}^1 \rho^{k}_{1}\odot \rho^{k}_{2}$, with $\rho_{i}^{0}=\frac{1}{2} \proj{+}$ and $\rho_i^{1}=\frac{1}{2^{n_1}} ((2^{n_1-1}-1)\proj{+}+ U_i \proj{+}U^{\dagger}_{i})$, for $i\in \{1,2\}$. The decoherence map for the case of two overlapping qubits is given by
$\E(\rho)=\rhosigma \odot \rho=\frac{1}{4}[3 \rho+\E_2 \circ \E_1(\rho)$],
where $\E_{j}(\rho)=\lambda_{20}\rho + \lambda_{21} U_j \rho U^{\dagger}_j$ and $\lambda_{20}=1-2^{2-n_1}$, $\lambda_{21}=2^{2-n_1}$. The corresponding reduced state is
$\rhosigma=4\sum_{k_1,k_2=0}^1 \rho^{k_1 k_2}_{1} \odot \rho^{k_1 k_2}_{2}$ with $\rho^{00}_{i}=\frac{1}{4} \proj{+}=\rho^{10}=\rho^{01}$ and $\rho^{11}_{i}=\frac{1}{2^{n_1}} ((2^{n_1-2}-1)\proj{+}+ U_i \proj{+}U_i ^{\dagger})$, for $i\in \{1,2\}$.
In the case where all qubits are overlapping, i.e. the gates act on the same bath qubits, we obtain $\E(\rho)=\rhosigma \odot \rho=2^{-n_1}[(2^{n_1}-1) \rho+U_1 U_2 \rho U^{\dagger}_1 U^{\dagger}_2]$. The reduced state $\rhosigma$ is then the single qubit reduced state of the state $U^{n_1+1}(\phi_1) U^{n_1+1}(\phi_2)\ket{+}^{\otimes n_1+1}=U^{n_1+1}(\phi_1+\phi_2)\ket{+}^{\otimes n_1+1}$.

\begin{figure}[h!]
\label{Fig1}
\includegraphics[width=1\columnwidth]{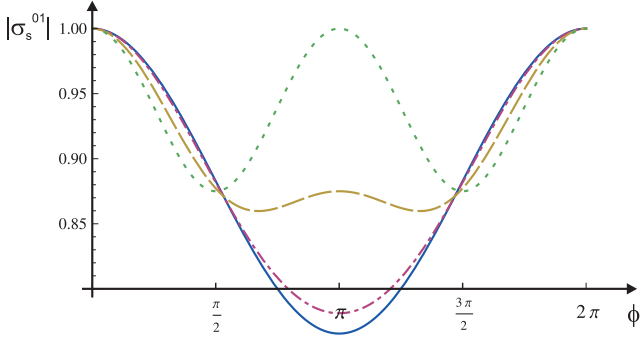}
\caption{For a 5--qubit phase gate the absolute value of $\bra{0}\rhosigma \ket{1}$ is plotted for different values of m. The continuous line shows the Markovian case where the two phase gates have no overlapping qubit, i.e. $m=0$. The dashed-dotted line corredsponds to one overlapping qubit ($m=1$) and the dashed line to three overlapping qubits ($m=3$). The dotted line refers to the extreme non-Markovian case with four overlapping qubits ($m=4$).
   }
\end{figure}

It can be seen from the equation above that for a fixed order of the phase gates, $n_1$, the coefficient in front of $\rho$ gets larger as $m$ increases. Thus, the system gets less disturbed if many of the bath qubits are overlapping. In order to investigate how the coherence for the single system qubit is affected by the number of coupled qubits, $m$, we plot the absolute value of the off-diagonal element of $\rhosigma$ for different values of $m$. Fig \ref{Fig1} shows the value of $\vert \rhosigma^{10} \vert$ for $5$-qubit phase gates for different number of coupled qubits $m=0,1,3,4$ and for $\phi_1=\phi_2=\phi$.
The extreme cases where no bath qubits are overlapping, $m=0$, and the case where all bath qubits are coupled, $m=4$, correspond to the Markovian and extreme non-Markovian case as already studied in Sec \ref{SecMarkovian}.
One main difference between those two cases is that the the non-Markovian case shows an oscillating behavior in contrast to the Markovian one.
When both phases are $\pi$, the value of $\vert \rhosigma^{10} \vert$ is maximal for the non-Markovian case, since there the map acts as the identity map. However, it is easy to see that in the Markovian case the value of $\vert \rhosigma^{10} \vert$ is minimal then. From the cases $m=1$ and $m=3$ one can infer the transition between the extreme cases. The $m=1$ case looks very like the Markovian case, whereas the $m=3$ case shows  already an oscillating behavior.
For small phases the Markovian case is less damped than the non-Markovian one.
This can be explained as follows. In the Markovian case the value of  $\vert \rhosigma^{10} \vert$ is multiplied $k$ times with itself, where $k$ denotes the number of phase gates applied to the qubit. Since $k$ equals only $2$ in our case, the value remains relatively large since $\vert \rhosigma^{10} \vert\approx1$.
The value of $\vert \rhosigma^{10} \vert$ for the non-Markovian case is decreasing, because we add the two phases of the two applied gates and therefore the value of the cosine appearing in the formula gets smaller. This effect is for small phases larger than in the Markovian scenario.

\subsubsection{Two system qubits}
In this subsection we consider two system qubits which are both interacting via a $(n_1+1)$--qubit phase gate with the environment.
Similarly to the single qubit case we investigate here how the degree of dependency in the baths influences the evolution of the system. Here, we will also study how the entanglement between the system qubits is affected by the different couplings within the bath.

The decoherence map for the case where two independent $(n_1+1)$-qubit phase gates are acting on a two qubit initial state ($m=0$) has the following form:
$\E(\rho)=\rhosigma \odot \rho= \E_2 \otimes \E_1(\rho)$ with
$\E_{j}(\sigma)= (1-2^{-n_1}) \sigma + 2^{-n_1}U_j \sigma U^{\dagger}_j)$.
The reduced state is given by $\rhosigma=\rho_1 \otimes \rho_2$ with $\rho_i$ being the single qubit reduced state of $U^{n_1+1}(\phi_i) \ket{+}^{\otimes n_1+1}$ (see Sec \ref{redStates}). For $m=1$, i.e. one bath qubit is overlapping, the decoherence map is given by
$\E(\rho)=\rhosigma \odot \rho= \frac{1}{2} \rho+\frac{1}{2}\E_2 \otimes \E_1(\rho)$ with $\E_{j}(\rho)=(1-2^{1-n_1})\sigma +2^{1-n_1} U_j \sigma U^{\dagger}_j$. The corresponding reduced state is
$\rhosigma=2\sum_{k=0}^1 \rho^{k}_{1}\otimes \rho^{k}_{2}$, with $\rho^{0}=\frac{1}{2} \proj{+}$ and $\rho^{1}=\frac{1}{2^{n_1}} ((2^{n_1-1}-1)\proj{+}+ U_i \proj{+}U^{\dagger}_{i})$. The case where two qubits are overlapping, i.e. $m=2$, is described by $\E(\rho)=\rhosigma \odot \rho=\frac{1}{4}[3 \rho+ \E_2 \otimes \E_1(\rho)]$ with $\E_{j}(\sigma)=(1-2^{2-n_1})\sigma + 2^{2-n_1} U_j \sigma U^{\dagger}_j)$. The corresponding reduced state is given by
$\rhosigma=4\sum_{k_1,k_2=0}^1 \rho^{k_1 k_2}_{1} \otimes \rho^{k_1 k_2}_{2}$ with $\rho^{00}_{j}=\frac{1}{4} \proj{+}=\rho^{10}=\rho^{01}$ and $\rho^{11}_{j}=\frac{1}{2^{n_1}} ((2^{n_1-2}-1)\proj{+}+ U_j\proj{+}U_j^{\dagger})$. In the extreme non-Markovian case where both system qubits interact with the same $n_1$ bath qubits, the map can be written as $\E(\rho)=(1-2^{-n_1})\rho+ 2^{-n_1}U_1 U_2 \rho U^{\dagger}_1 U^{\dagger}_2$
with the reduced state given by $\rhosigma=2^{n_1}\sum_{k_1=0}^1 \rho^{k_1 k_2 \ldots k_{n_1}}_{1} \otimes \rho^{k_1 k_2 \ldots k_{n_1}}_{2}=(1-2^{-n_1})\proj{++}+2^{-n_1}U_1 U_2 \proj{++} U^{\dagger}_1 U^{\dagger}_2)$ with $\rho_{j}^{k_1 k_2 \ldots k_{n_1}}=\frac{1}{2^{n_1}}\proj{+}$ $\forall$ $k_1 \cdot k_2 \cdot \ldots k_{n_1}=0$ and $\rho_{j}^{k_1 k_2 \ldots k_{n_1}}=\frac{1}{2^{n_1}}U_j\proj{+} U^{\dagger}_j$ $\forall$ $k_1 \cdot k_2 \cdot \ldots k_{n_1}=1$. Similar to the single qubit case investigated before, we see that the factor in front of $\rho$ increases as $m$ increases. Thus, the system gets less distorted the more bath qubits are overlapping.

Let us now study the evolution of the entanglement between the system qubits for several different coupling schemes. As examples we consider $n_1=2$, that is the system qubits are interacting with the environment via $3$--qubit phase gates. We compare three different types of coupling scenarios: 1) the case where each gate acts independently on one of the system qubits ($m=0$); 2) the case where one bath qubits overlaps ($m=1=n_1-1$) and; 3) the case where two bath qubits are overlapping ($m=2=n_1$).
We compute the entanglement of formation of the output state $\E(\rho)$ for the different couplings. The entanglement of formation is numerically optimized with respect to a maximally entangled input state.
\begin{figure}[h!]
\label{Fig2}
\includegraphics[width=0.65\columnwidth]{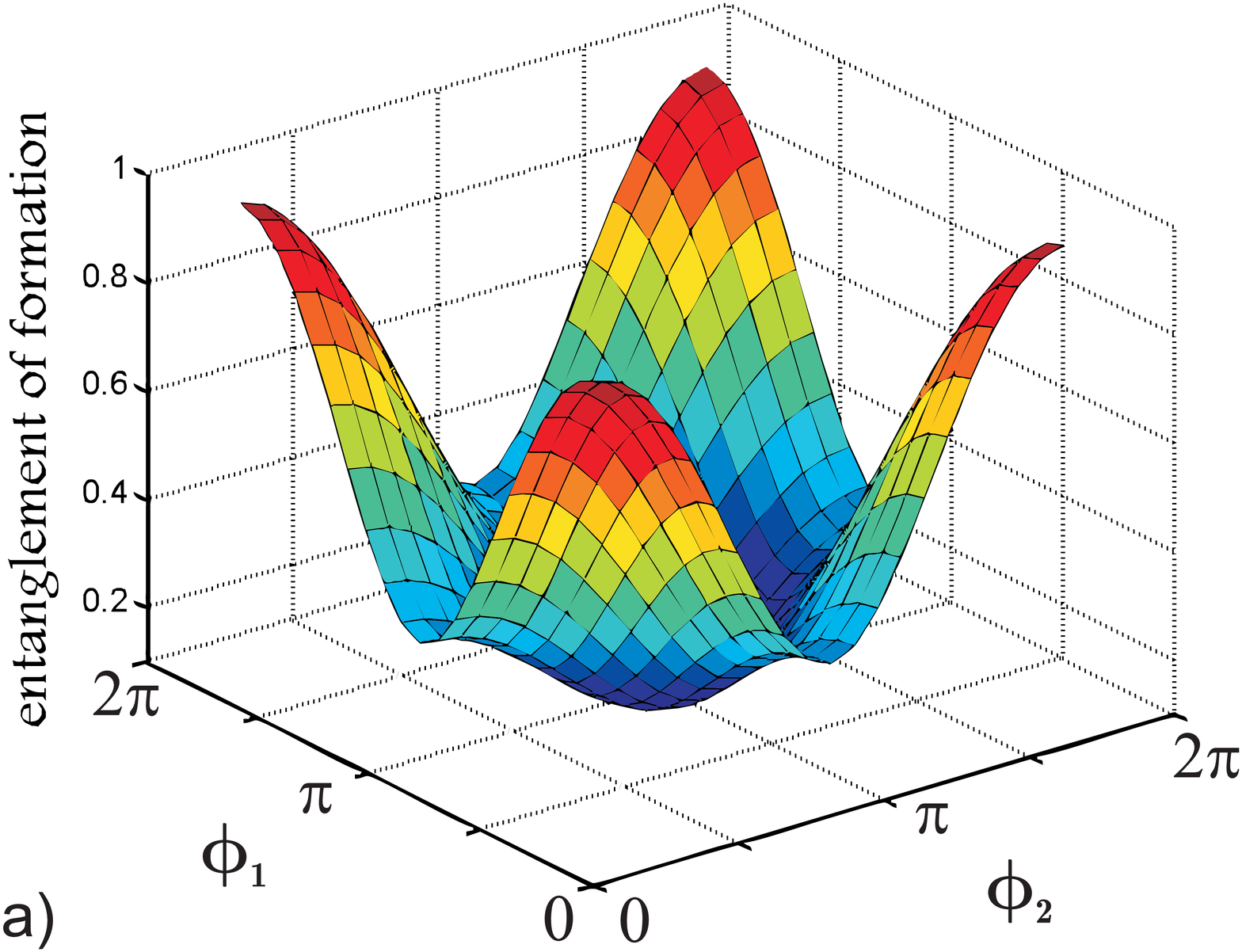}
\includegraphics[width=0.65\columnwidth]{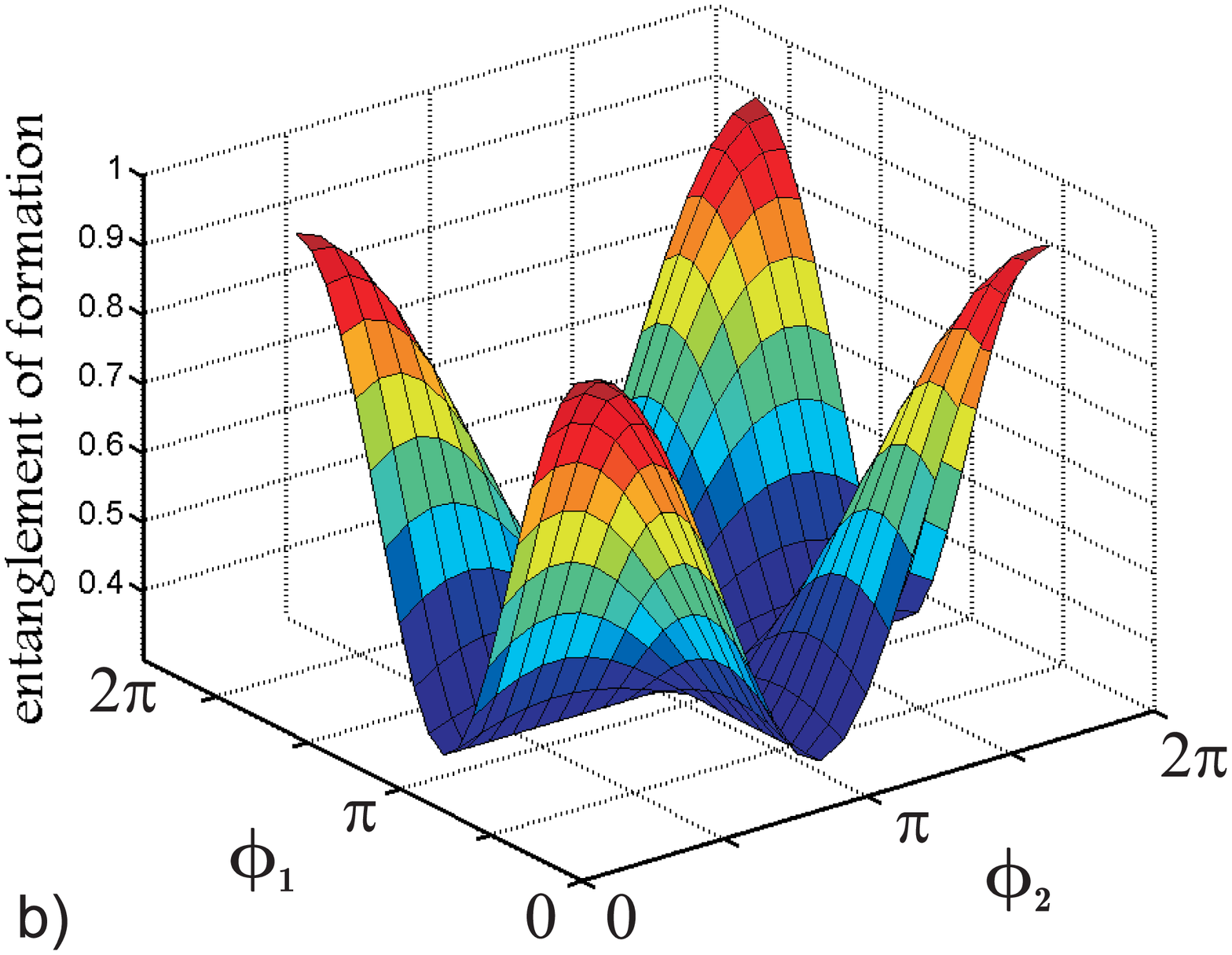}
\includegraphics[width=0.65\columnwidth]{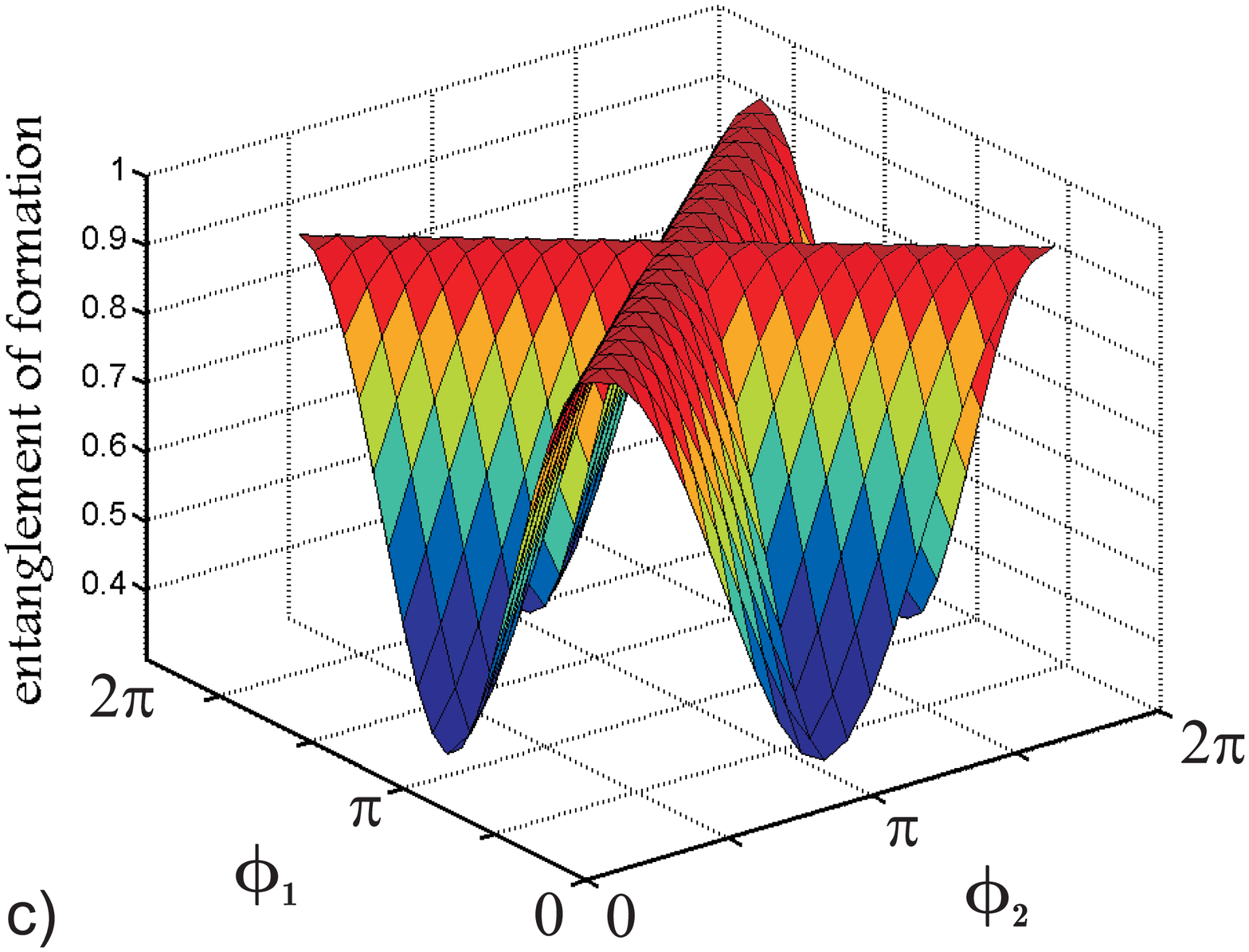}
\caption{entanglement of formation optimized with respect to a maximally entangled input state for three different coupling scenarios a$)$ two three qubit phase gates $U_{1b_1 b_2}U_{2b_3 b_4}$ acting independently on the two system qubits $1$ and $2$  b$)$ one bath qubit is overlapping within the phase gates $U_{1b_1 b_2}U_{2b_2 b_3}$   c$)$ the phase gates $U_{1b_1 b_2}U_{2b_1 b_2}$ have one overlapping edge}
\end{figure}
Fig. \ref{Fig2} shows the entanglement of formation of the optimized input state as a function of the two phases, $\phi_1$ and $\phi_2$, of the three--qubit phase gates. In case 1), where $m=0$, the minimum at $\phi_1=\phi_2=\pi$, can be explained as follows. As explained in Sec \ref{SecMarkovian} considering a single qubit interacting with its bath via a pure $(n_1+1)$--qubit phase gate with phase $\phi$ generates a maximally entangled state only if $\phi=\pi$ and $n_1=1$.
For larger values of $n_1$, the entanglement is still optimal, however it is no longer maximal, for $\phi=\pi$. Thus, if both system qubits are interacting with their independent baths, the entanglement between them is minimized if both phases are $\pi$ due to entanglement monogamy.

In the second case, it can be seen in Fig \ref{Fig2} that the entanglement is minimized if one of the phases is $\pi$. This fact might be explained by the following observation. The map which is applied to the input state is given by $\E(\rho)=\frac{1}{2} \rho+\frac{1}{2} \E_1\otimes\E_2(\rho)$ with $\E_j(\sigma)=\frac{1}{2} \sigma+\frac{1}{2} U_j \sigma U^{\dagger}_{j}$. As mentioned in Sec \ref{SubsecPurelyDephasing} we can interpret $\E_j(\sigma)$ as the map corresponding to the interaction of the system qubit with a single bath qubit via a two--qubit phase gate, $U$, i.e. $\E_j(\sigma)=\tr_B (U \proj{+}\otimes \sigma U^\dagger)$. Using this fact it is now easier to see why there is a minimum if one of the phases is $\pi$. Similar to the reasoning above the two qubit phase gates have its maximal entangling capability between the system and the bath for $\phi=\pi$ and therefore the entanglement between the system qubits after applying the map $\E_1\otimes\E_2$ is minimized due to entanglement monogamy.

In the third case, where $m=n_1=2$, we obtain a maximally entangled state whenever $\phi_1=\phi_2$ or $\phi_1+\phi_2=2\pi$. This can be easily explained by noting that the output state is of the form ${\cal E}(\rho)=\alpha \rho+\beta U_1\otimes U_2\rho U_1^\dagger \otimes U_2^\dagger$, for some values of $\alpha$ and $\beta$. Now, if $\phi_1=\phi_2$, which implies that $U_1=U_2$ the singlet state, $\ket{\Psi^-}=1/\sqrt{2}(\ket{01}+\ket{10})$ is left invariant under this map. Therefore, the output state is maximally entangled. In the other case, where $\phi_1+\phi_2=2\pi$, which implies that $U_2=U_1^\dagger$, $\ket{\Phi^+}$ is left invariant. Thus in both cases the optimal entanglement of formation, optimized with respect to the maximally entangled input state, is maximal.

\subsubsection{n system qubits}

We consider the evolution of $N_S$ system qubits, where each system qubit, $s_l$ is interacting with $n_1$ bath qubits via a pure $(n_1+1)$--phase gate with phase $\phi_l$. The degree of bath--correlation is quantified by the number of overlapping bath qubits, $m$, which we again denote by $b_1,\ldots, b_m$. Those bath qubits are interacting with all system qubits and no other bath qubit is interacting with more than one system qubit (see conditions (a) and (b)). The evolution of the system is governed by the completely positive map given in Eq (\ref{locmaps}). Let us now explicitly write down the completely positive map for the three cases: 1) $m=n_1$, 2) $m=n_1-1$, and 3) $m=0$. 1) For $m=n_1$ all system qubits are coupled to the same $m$ bath qubits via a $(n_1+1)$--phase gate, i.e. the baths of all system qubits are maximally overlapping. In this case we have ${\cal E}(\rho)=\alpha \rho+\beta\bigotimes_i U_i \rho U_i^\dagger$ with $\alpha=1-2^{-n_1}$ and $\beta=2^{-n_1}$. 2) For $m=n_1-1$ all system qubits are connected to the bath qubits $\{b_1,\ldots b_{n_1-1}\}$ and one additional bath qubit, which is not connected to any other qubit. In this case we have ${\cal E}(\rho)=\alpha\rho+\beta (\bigotimes {\cal E}_i ) (\rho)$ with $\alpha=1-2^{-n_1+1}$ and $\beta=2^{-n_1+1}$. Here, ${\cal E}_l(\sigma)=\alpha_l \sigma+ \beta_l U_l \sigma U_l^\dagger$ with $\alpha_l=\beta_l=\frac{1}{2}$. 3) For $m=0$ each system qubit is interacting with its own bath. In this case the evolution if governed by the completely positive map ${\cal E}(\rho)=(\bigotimes {\cal E}_l ) (\rho)$, where ${\cal E}_l(\sigma)=\alpha_l \sigma+ \beta_l U_l \sigma U_l^\dagger$ with $\alpha_l=1-2^{-n_l}$ and $\beta_l=2^{-n_l}$.
As can be seen above (see also Eq (\ref{locmapszwei})) the decoherence map is always of the form ${\cal E}(\rho)=\alpha \rho +\beta \tilde{{\cal E}}(\rho)$, where $\tilde{{\cal E}}(\rho)$ denotes a local map. Since the coefficient $\alpha$ is given by $1-2^{-m}$, one might expect that the system gets less disturbed the larger $m$ gets, that is the stronger the coupling among the baths is. However, as we will see below, this is not the case in general. For instance, the entanglement of a state can be decreased faster when the system qubits are interacting with a maximally dependent bath than it is if each system qubit is interacting with its own bath.

As example we consider the input state $\ket{\Psi}=\sqrt{p}\ket{0}^{\otimes N_S}+\sqrt{1-p}\ket{1}^{\otimes N_S}$ with $p>0$.
We assume that all interactions with the environment are equivalent, i.e. $\phi_l=\phi$, $\forall l$.
It is straightforward to compute the output state for the different couplings mentioned above. Let us compare here the two extreme cases, $m=0$ and $m=n_1$. In order to quantify the entanglement of the output state we consider here the bipartite splitting of one system qubit versus the remaining $N_S-1$ qubits. Since the input state and the interaction is symmetric with respect to particle exchange, we consider, without loss generality the entanglement, $E$, between qubit $1$ and the rest, which we measure with the $\log$--negativity \cite{WeVi08}. It can be easily shown that for the case, where each system qubit is interacting with its own bath we find $E_{indep}=\log_{2}(1+2\vert \sqrt{p}\sqrt{1-p}\sqrt{(1-2^{-n_1}+e^{i\phi} 2^{-n_1})^{N_S}}\vert$) whereas in the extreme non--Markovian case we have $E_{dep}=\log_{2}(1+2\vert \sqrt{p}\sqrt{1-p}\sqrt{1-2^{-n_1}+e^{i N_S \phi} 2^{-n_1}}\vert$).
The coupling of each system qubit to its individual bath destroys less entanglement than a correlated bath if $E_{indep}>E_{dep}$. We are going to show next that such a scenario is possible for an arbitrary number of system qubits, $N_S$. Considering three--qubit phase gates, $n_1=2$, with $\phi=\frac{\pi}{N_S}$ we find $E_{indep}>E_{dep}$ iff $[\frac{1}{4}(\frac{10}{16}+\frac{6}{16}\cos(\pi/N_S))]^{\frac{N_S}{2}}<1$. Since the left hand side of this inequality can be easily upper bounded by $[\frac{1}{4}]^{\frac{N_S}{2}}$ this shows that in general it is not true that the entanglement is less distorted if the degree of bath--correlation is larger.

\subsection{Comparison order of gates}
\label{SecOrder}

In this section we investigate the effect of the order $n_l$ of the phase gates on the evolution of the system. We will fix the number of overlapping bath qubits $m$. The influence of $n_l$ can be seen by looking at the decoherence map presented in Eq. (\ref{locmapszwei}), i.e. ${\cal E}_m^{n_1,\ldots, n_{N_S}}(\rho)=\frac{1}{2^{N_B}}[(2^{N_B}-2^{N_B-m}) \rho+2^{N_B-m}\bigotimes_{l=1}^{N_S}{\cal E}_l(\rho)], \label{locmaps2}$ where $\E_{l}(\sigma)=((1-2^{m-n_l})\sigma+2^{m-n_l}U_l(\phi_l)\sigma U_l(\phi_l)^\dagger)$. The order of the phase gates, $n_l$, does not affect the coefficient in front of $\rho$, but changes only the weights in the maps ${\cal E}_l$. The larger $n_l$  the larger is the coefficient in front of $\sigma$.
Therefore the initial state is less altered by phase gates of higher order. This coincides with the fact that the probability of applying a phase gate on the system qubit gets smaller the larger the order of the phase gate is.  For example, if the system interacts with the bath via a $5$--qubit phase gate, all four bath qubits have to be in the state $\ket{1}$ in order to apply the phase gate to the system qubit; for a $2$--qubit gate, however, just one bath qubit has to be in the state $\ket{1}$.

\section{Generating LMESs}
\label{SecLMEgen}

In this section we will consider the complementary process to the one investigated so far. We will use the coupling of the system qubits to the bath in order to  generate a state of interest. This can then be used to either prepare the state using dissipation, or a unitary evolution.

We call a complete set of commuting unitary observables,
$\{U_i\}_{i=0}^n$ generalized stabilizer if the eigenvalues of
$U_i$ are $\pm 1$ and both, $1$ and $-1$ are $2^{n-1}$--fold
degenerate \footnote{Note that for stabilizer states, the stabilizers $U_i$ would all be elements of the Pauli group. In general however, there is no reason to restrict to those stabilizers.}. Furthermore, we call a  $n$ partite state, $\ket{\Psi}$
a generalized stabilizer state, if there exists a generalized
stabilizer, $\{U_i\}_{i=0}^n$, such that $U_i\ket{\Phi}=\ket{\Phi}$
$\forall i$ iff $\ket{\Phi}=\ket{\Psi}$. Examples of generalized stabilizer states are all LMESs. For an arbitrary $n$--qubit LMES, $\ket{\Psi}=U^\Psi_{ph} \ket{+}^{\otimes n}$, the generalized stabilizer are of the form
$U_k= U^\Psi_{ph} X_k
(U^\Psi_{ph})^\dagger$ where $\x$ denotes here and in the following the Pauli operator $\sigma_x$. It is easy to verify that $U_k\ket{\Phi}=\ket{\Phi}$ $\forall k$
iff $\ket{\Phi}=\ket{\Psi}$. Depending only on the phases $\alpha_{\bf i}$, which define
the LME, $\ket{\Psi}$, the generators of the generalized stabilizer can be
quasi--local, i.e. act non trivially on a small set of
(neighboring) qubits. In this case, the methods
developed in \cite{VeWoCi08} can be employed to derive a
quasi--local dissipative process for which the unique stationary
state is $\ket{\Psi}$. Apart form that, one can also easily
construct a frustration free Hamiltonian for which the unique
ground--state is $\ket{\Psi}$ \cite{KrKr08}.

In the following we show how generalized stabilizers can be
prepared by a unitary evolution. Therefore, we use $n$ auxiliary
qubits which are all prepared in the state
$\ket{+}$. The unitary operation is
composed out of control operations, where the auxiliary systems act
as control qubits. The following theorem shows that a generalized
stabilizer state can always be prepared by applying this unitary to
an arbitrary input state of the system.

\begin{theorem} \label{ThGenLME}

If $\ket{\Psi}$ is a $n$--qubit generalized stabilizer state then
there exists a complete set of commuting unitary observables,
$\{U_i\}_{i=0}^n$ (the generalized stabilizer) and a set
$\{V_i\}_{i=0}^n$ with $\{V_i,U_i\}=0$ such that for all $n$--qubit
states $\ket{\Phi}$ there exists a $n$--qubit state $\ket{\phi}$
such that \bea \ket{\Psi}_s\ket{\phi}_a=U^c_1 U^c_2 \ldots U^c_n
\ket{\Phi}_s\ket{+}_a^{\otimes n}.\eea

The control operations $U_i^c$ are of the form
$U^c_i=\bar{U}_i\tilde{U}_i$, where $\tilde{U}_i=\one \otimes
\ket{0}_{a_i}\bra{0}+U_i \otimes \ket{1}_{a_i}\bra{1}$ and
$\bar{U}_i=\one \otimes \ket{+}_{a_i}\bra{+}+V_i \otimes
\ket{-}_{a_i}\bra{-}$.

\end{theorem}

Thus, in order to prepare a generalized stabilizer state, one only
has to prepare the auxiliary systems in the product state
$\ket{+}^{\otimes n}$ and apply certain control gates. Depending on
the properties of the state $\ket{\Psi}$ these gates might be
quasi--local, i.e. act only on a few (neighboring) qubits. Note
that this theorem can also be used to derive a process using either
dissipation, or a completely positive map in order to prepare the
state, $\ket{\Psi}$ \cite{VeWoCi08}.

\begin{proof}

First of all, note that the set $\{U_i\}$ does not only define the
state $\ket{\Psi}$, but a whole set, the common eigenbasis of the
commuting observables. Each element of this basis is uniquely
defined by its eigenvalues. We use the notation
$\ket{\Psi_{i_1,\ldots,i_n}}$ for the basis elements with
eigenvalues $((-1)^{i_1},\ldots (-1)^{i_n})$, with $i_k\in
\{0,1\}$. Let us denote by $S_0^i\equiv {\mbox
span}\{\ket{\Psi_{k_1,\ldots, k_i=0,\ldots,k_n}}_{k_l\in \{0,1\}}$
($S_1^i\equiv {\mbox span}\{\ket{\Psi_{k_1,\ldots,
k_i=1,\ldots,k_n}}_{k_l\in \{0,1\}}$) the range of $\one+U_i$
($\one-U_i$) respectively. That is any state which is an eigenstate of $U_i$
with eigenvalue $1$ ($-1$) is within $S_0^i$ ($S_1^i$) respectively
and vice versa. Note that both subspaces have dimension $2^{n-1}$.
We define $V_i$ as the unitary which exchanges $S_0^i$ with
$S_1^i$. To be more precise, $V_i=\sum_{k_1,\ldots k_n\in \{0,1\}}
\ket{\Psi_{k_1,\ldots, k_i=0,\ldots,k_n}}\bra{\Psi_{k_1,\ldots,
k_i=1,\ldots,k_n}}+ h.c.$. Now, we decompose $\ket{\Phi}$ into a
state which is in $S_0^n$ and one which is in $S_1^n$, i.e.
$\ket{\Phi}=(\one+U_n)\ket{\Phi}+(\one-U_n)\ket{\Phi}\equiv
\ket{\Phi_n^0}+\ket{\Phi_n^1}$. Applying $\tilde{U}_n$ to the state
$\ket{\Phi} \ket{+}_{a_n}$ leads to
$\tilde{U}_n\ket{\Phi}\ket{+}_{a_n}=\ket{\Phi_n^0}(\ket{0}+\ket{1})_{a_n}+\ket{\Phi_n^1}(\ket{0}-\ket{1})_{a_n}$.
Applying now $\bar{U}_n$ leads to $U_n^c
\ket{\Phi}\ket{+}_{a_n}=\ket{\Phi_n^0}\ket{+}_{a_n}+V_n\ket{\Phi_n^1}\ket{-}_{a_n}$,
where both, $\ket{\Phi_n^0}$ and $V_n\ket{\Phi_n^1}$ are elements
of $S_0^n$. Continuing in this way, we find that all the terms
which occur are elements of $S_0^1\bigcap S_0^2 \bigcap \ldots
\bigcap S_0^n=\{\ket{\Psi}\}$. This is due to the fact that the
operators $V_i$ only change a state in $S_1^i$ to a state in
$S_0^i$ and vice versa, but if the resulting state is an
eigenstate of $U_k$ with eigenvalue $1$ ($-1$), it remains an
eigenstate of $U_k$ with the same eigenvalue.

\end{proof}

Theorem \ref{ThGenLME} can also be used to show how an
arbitrary state can be constructed using a generalized stabilizer state as a recourse.
This can be seen as follows. Due to Theorem \ref{ThGenLME} we have that for any $n$--qubit state, $\ket{\Phi}$ there exists a $n$--qubit state $\ket{\phi}$
such that $\ket{\Psi}_s\ket{\phi}_a=U^c_1 U^c_2 \ldots U^c_n
\ket{\Phi}_s\ket{+}_a^{\otimes n}$. Thus, we have $\ket{\phi}_a=\bra{\Psi}_sU^c_1 U^c_2 \ldots U^c_n
\ket{\Phi}_s\ket{+}_a^{\otimes n}$. Note that \bea U^c_1 U^c_2 \ldots U^c_n
\ket{\Phi}_s\ket{+}_a^{\otimes n}=\\ \nonumber \sum_{i_1,\ldots
i_n} X_1^{(i_1)}\cdots X_n^{(i_n)} \ket{\Phi} H^{\otimes
n}\ket{i_1,\ldots i_n}.\eea Here, $H$ denotes the Hadamard
transformation, and $X_j^{(0)}=\one+U_j$,
and $X_j^{(1)}=V_j(\one-U_j)=(\one+U_j)V_j$. Since $U_j$ is
hermitian, we have that $\bra{\Psi} X_1^{(i_1)}\cdots
X_n^{(i_n)}=\bra{\Psi_{i_1,\ldots,i_n}}$. Denoting by $U_{\Psi}$
the unitary which transforms the computational basis into the basis
$\ket{\Psi_{i_1,\ldots,i_n}}$, i.e. $U_{\Psi}=\sum_{i_1,\ldots i_n}
\ket{\Psi_{i_1,\ldots,i_n}}\bra{i_1,\ldots,i_n}$ we have \bea \label{Eqphi}
\ket{\phi}=H^{\otimes n} U_\Psi^\dagger \ket{\Phi}.\eea

Thus, an arbitrary state $\ket{\Phi}$ can be prepared by applying the controlled operations $(U_i^c)^\dagger$ to the state $\ket{\Psi}\ket{\phi}$, where $\ket{\phi}$ is given in Eq. (\ref{Eqphi}). For any LMES we have
$\ket{\Psi}=U_\Psi\ket{0,\ldots ,0}$. As before, one can define a phase--gate, $U_{ph}^\Psi$,
which is diagonal in the computation basis, such that, $\ket{\Psi}=
U_{ph}^\Psi \ket{+}^{\otimes n}$, and, more generally,
$U_\Psi=U_{ph}^\Psi H^{\otimes n}$. In this case we would have
$\ket{\phi}=(U_{ph}^\Psi)^\dagger \ket{\Phi}$. Choosing $\ket{\Phi}=\ket{+}^{\otimes n}$ we obtain $\ket{\phi}=\ket{\Psi^\ast}$, where $\ket{\Psi^\ast}$ denotes the complex conjugation of $\ket{\Psi}$ in the computational basis. Hence we have,
\bea \ket{\Psi}\ket{\Psi^\ast}= U^c_1
U^c_2 \ldots U^c_n \ket{+}_s^{\otimes n}\ket{+}_a^{\otimes n}.\eea
It should be noted here that for an arbitrary number of qubits, $n$, there exists a $n$--qubit LMES, $\ket{\Psi}$, such that $\ket{\Psi}$ and $\ket{\Psi^\ast}$ are neither LU--equivalent, nor can be transformed into each other by more general operations, namely local operations and classical communication \cite{Kr10}. For Graph states,
into which any stabilizer state can be transformed by local unitary
operations, and also for certain LMESs the phase gate is
hermitian, which implies that $\ket{\Psi}=\ket{\Psi^\ast}$. Therefore, we find for those states \bea \ket{\Psi}\ket{\Psi}= U^c_1
U^c_2 \ldots U^c_n \ket{+}_s^{\otimes n}\ket{+}_a^{\otimes n}.\eea
Note that this is a new method to generate two copies of an
entangled state, where both of them are generated in one step. That is, the states are not generated independently. One might compute the entanglement of formation
using this procedure and compare it to twice the entanglement of
formation of a single copy of $\ket{\Psi}$.\\

\section{Conclusion and outlook}

We investigated the effect of decoherence induced by many--body interactions. In our decoherence model we considered multi-qubit Ising interactions between the system and bath qubits. These interactions could result from multipartite collisions. The corresponding unitary was modeled as a multipartite phase gate. Due to the large variety of different interaction patterns we focused here on some relevant aspects of decoherence. We mainly considered purely dephasing maps, where each system qubit is coupled to another system qubit only via the environment, i.e. there is no collision between system qubits. We showed that all purely dephasing maps are separable. In order to analyze the effect of correlations within the bath on the evolution of the system we considered those situations where $m$ bath qubits are connected to all system qubits, which can also interact with other, non--overlapping bath qubits. We derived a simple expression for those maps and showed how the effect of decoherence depends on the number of commonly shared bath qubits, $m$. This enabled us to compare all the cases, from the one where each system is coupled to its own environment to the one where all system qubits are coupled to the same environment. Furthermore, we compared the Markovian to the Non--Markovian processes. We also showed how the evolution of the system is influenced by the order of applied phase gates. Note that these results can also be generalized to higher dimensional systems. The coupling to the environment can not only destroy the coherence and entanglement in the system, but can also be used to accomplish certain tasks \cite{VeWoCi08}. We showed here how the bath qubits can be used to prepare a desired state. In particular, we derived a unitary operation, acting on the bath and system qubits which can be employed to generate certain multipartite states. This investigation also lead to a protocol which prepares two copies of certain multipartite states in a single step.

An other interesting question in this context is how the entanglement established between the system and bath qubits is related to the induced decoherence. For instance, the dephasing map ${\cal E}(\rho)=\rhosigma \bigodot \rho$ is a completely dephasing map, i.e. ${\cal E}(\rho)=\rho_D$, where $\rho_D$ is a diagonal matrix whose diagonal is the same as the one of $\rho$, for any state $\rho$, iff $\rhosigma=\one$. That is the map destroys all the coherence in the system iff the corresponding LMES is maximally entangled between the system and the bath qubits. A simple example would be the state $\ket{\Psi}_{SB}=U_{SB} \ket{+}^{\otimes 2n}=\ket{\Phi^+}_{SB}^{\otimes n}$. Here, $\ket{\Phi^+}$ denotes a maximally entangled state shared between a bath and a system qubit. Obviously, for states, $\rho$, whose diagonal elements are all equal, such a map would completely depolarize the state. Examples of such states are arbitrary convex combination of product LMES (all in the computational basis).

An other interesting topic is the characterization of decoherence free subsets, i.e. the characterization of those states which are left unchanged under certain dephasing map. For instance, if we consider $N_S$ system qubits, which are all coupled to the same bath composed of $m$ qubits, i.e. $n_l=m+1$ for all $l$ we have (see Eq. (\ref{locmapszwei})) ${\cal E}(\rho)=\alpha \rho+ \beta U_1 \otimes \ldots \otimes U_{N_S} \rho U_1^\dagger \otimes \ldots \otimes U_{N_S}^\dagger$. Thus, a state $\rho$ is left invariant under this coupling iff $U_1 \otimes \ldots \otimes U_{N_S} \rho U_1^\dagger \otimes \ldots \otimes U_{N_S}^\dagger=\rho$. Choosing the phases such that all local unitaries coincide, we see that the state $\rho$ will be invariant iff $U^{\otimes N_S}\rho (U^{\otimes N_S})^\dagger=\rho$. States which fulfill the above condition for any unitray $U$ are called super--singlets \cite{Cab03}. Since we consider here only dephasing maps, the state must only be invariant under local phase gates, which implies that the decoherence--free subset contains more states than the super--singlets. For instance, in the case of two qubits, the so--called Werner states, which are of the form $\rho=(1-p)/4 \one + p \proj{\Psi^-}$, where $1\geq p\geq 0$ and $\ket{\Psi^-}$ denotes the singlet state, characterize all states invariant under $U\otimes U$, for any unitary $U$ \cite{Wer89}. However, the decoherence free subspace of the map described above contains any state of the form $\rho=\alpha \proj{00}+\beta \proj{11}+ \gamma \proj{\Psi^+}+\delta \proj{\Psi^-}$, for $\alpha,\beta,\gamma,\delta\geq 0$.

\section{Acknowledgment}
B.K. would like to thank Peter Zoller for fruitful discussions and T.C. and B.K. are grateful to Wolfgang D\"ur for helpful discussions. This work has been supported by the FWF (Elise Richter Program, SFB FOQUS).

\end{document}